\def\imat  {i}
\def\be{\begin{equation}}
\def\ee{\end{equation}}
\def\bea{\begin{eqnarray}}
\def\eea{\end{eqnarray}}
\def\la{\langle}
\def\ra{\rangle}

\documentclass[aps,prb,twocolumn,amsmath,amssymb,superscriptaddress]{revtex4}
\usepackage{epsfig}
\usepackage{bm}

\begin{document}

\title{Uncollapsing the wavefunction by undoing quantum measurements}
\author{Andrew N. Jordan}
\affiliation{Department of Physics and Astronomy, University of
Rochester, Rochester, New York 14627, USA }
\author{Alexander N. Korotkov}
\affiliation{ Department of Electrical Engineering, University of
California, Riverside, CA 92521, USA}
\date{\today}
\begin{abstract}
We review and expand on recent advances in theory and experiments
concerning the problem of wavefunction uncollapse:  Given an unknown
state that has been disturbed by a generalized measurement, restore
the state to its initial configuration.   We describe how this is
probabilistically possible with a subsequent measurement that
involves erasing the information extracted about the state in the
first measurement.  The general theory of abstract measurements is
discussed, focusing on quantum information aspects of the problem,
in addition to investigating a variety of specific physical
situations and explicit measurement strategies.   Several systems
are considered in detail: the quantum double dot charge qubit
measured by a quantum point contact (with and without Hamiltonian
dynamics), the superconducting phase qubit monitored by a SQUID
detector, and an arbitrary number of entangled charge qubits.
Furthermore, uncollapse strategies for the quantum dot electron
spin qubit, and the optical polarization qubit are also reviewed.
For each of these systems the physics of the continuous measurement
process, the strategy required to ideally uncollapse the
wavefunction, as well as the statistical features associated with
the measurement is discussed.  We also summarize the recent
experimental realization of two of these systems, the phase qubit
and the polarization qubit.
\end{abstract}

\maketitle

\section{Introduction}
The irreversibility of quantum measurement is an axiomatic property
of textbook  quantum mechanics.\cite{Neumann}  In his famous article
{\it Law without law}, John Wheeler expresses the idea with poetic
flare:  ``We are dealing with [a quantum] event that makes itself
known by an irreversible act of amplification, by an indelible
record, an act of registration.''\cite{wheeler}  However, it has
been gradually recognized that the textbook treatment of an
instantaneous wavefunction collapse is really a very special case of
what is in general a dynamical process - continuous quantum
measurement.  Continuous measurements do not project the system
immediately into an eigenstate of the observable, but describe a
process whereby the collapse happens over a period of time.
\cite{Davies-Kraus-Holevo} The fact that continuous measurement is a
dynamical process with projective measurement as a special case,
leads us to ask whether the irreversibility of quantum measurement
is also a special case. The purpose of this paper is to review and
expand on recent developments in this area of research, showing that
it is possible to undo a quantum measurement, thereby uncollapsing
the wavefunction, and to describe this physics in detail for both
the abstract and concrete physical realizations.

This paper follows our earlier work on the
subject,\cite{us-undo,us-proceeding} as well as other papers
investigating similar questions.\cite{Ueda-99,others}  Wavefunction
uncollapse teaches us several things about the fundamentals of
quantum mechanics.  First, there is a notion that wavefunction
represents many possibilities, but that reality is created by
measurement.  The fact that the effects of measurement can be undone
suggests that this idea is flawed, or at least too simplistic. If
you create reality with quantum measurements, does undoing them
erase the reality you created?  Secondly, there is a wide-spread
belief that quantum measurement is nothing more than a decoherence
process.  This suggests that the superposition never really
collapses; it only appears to collapse. What actually happens,
according to this idea, is that all the information about the system
disperses into the environment: when a quantum system interacts with
a classical measuring device, it becomes irreversibly entangled with
all the particles that make up the measuring device and its
surroundings. The uncollapse of the wavefunction demonstrates that
decoherence theory cannot be the whole story, because a true
decoherence process is irreversible.\cite{gefter}   The perspective
we take in this paper  further advances the `quantum Bayesian' point
of view, where the quantum state is nothing more than a reflection
of our information about the system.  When we receive more
information about the system, the state changes or collapses not
because of any mysterious forces, but simply as a result of Bayesian
updating.

Our work is indirectly related to the `quantum eraser' of Scully and
Dr\"uhl.\cite{scully}    There, the which-path information of a
particle is encoded in the quantum state of an atom, resulting in a
destruction of interference fringe visibility.  On the other hand,
if the which-path information of the particle is erased, the
interference fringes are restored.  Both the Scully proposal and our
proposal erase information.  However, there is an important
difference.  In order for the uncollapsing procedure to work, we
have to erase the information that was already extracted
classically. In the `quantum eraser', only potentially extractable
information is erased.

While the first part of this paper deals with the abstract idea of
uncollapse, and formalizes its properties in terms of generalized
quantum measurements, a great deal of the paper deals with actual
measurement processes in specific physical contexts.  This brings to
mind the saying of Asher Peres: ``Quantum phenomena do not occur in
a Hilbert space.  They occur in a laboratory''.\cite{peres}
Following Peres' dictum, we discuss measurement
processes in a variety of solid state systems, where there has been
remarkable experimental progress in recent years.  Quantum coherence
has been demonstrated to occur in a controllable fashion in systems
such as semiconductor quantum dots and superconducting Josephson
junctions. We will discuss the physics of measurement in these
systems, as well as concrete strategies for uncollapsing the
wavefunction.  It should be stressed that two of these proposals
(a superconducting phase qubit and optical polarization qubit) have now been implemented in the laboratory, providing conclusive demonstration of wavefunction
uncollapse.\cite{Katz2}

The paper is organized as follows.  In Sec.~II we describe in detail
what we mean by the undoing of a quantum measurement (Sec.~II\,A),
and give a general treatment of the physics, using the formalism of
positive operator-valued measures (POVMs).  This is done both for
pure states (Sec.~II\,B) and mixed states (Sec.~II\,C).  In
Sec.~II\,D we discuss the interpretation of wavefunction uncollapse,
and what it tells us about quantum information.  Sec.~III begins
discussion of the physical implementation of this physics, with a
treatment of the double quantum dot qubit, monitored by a quantum
point contact.  The measurement dynamics is discussed in
\ref{charge-dynamics} and we discuss the uncollapsing strategy and
statistical predictions in Sec.~\ref{charge-uncollapse} for the case
of a qubit undergoing measurement dynamics only. The results are
discussed in \ref{charge-discussion} and compared with the abstract
results. In Sec.~\ref{charge-statistics} we discuss the statistical
features of the time required to wait until the wavefunction is
uncollapsed. We generalize to the case of the finite-Hamiltonian
qubit undergoing the uncollapse process in Sec.~\ref{charge-evolve},
and derive results for the success probability in that case.  We
switch to a new physical system in Sec.~IV, the superconducting
phase qubit. In Sec.\ IV\,A we discuss how the measurement process
works for the phase qubit, and in Sec.\ IV\,B we discuss the
uncollapsing strategy, corresponding success probability, and the
recent experimental realization, which demonstrated the
uncollapsing. In Sec.~V, we generalize to the case of many charge
qubits, and discuss an explicit procedure to undo any generalized
measurement.   We discuss recent developments in the theory and
experiments of measurement reversal in Sec. VI and conclude in
Sec.~VII.

\section{General theory of uncollapsing}
\label{general-theory}

\subsection{Preliminary discussion}\label{PrDis}

    Our goal is to restore an initial quantum state disturbed by measurement.
However, it is important to discuss what exactly we mean by that.
For example, if we start with a {\it known} pure state
$|\psi_{in}\rangle$ and perform a textbook projective measurement,
then it is trivial to restore the initial state: since we also know
the post-measurement wavefunction $|\psi_m\rangle$, we just need to
apply a unitary operation which transfers $|\psi_m\rangle$ into
$|\psi_{in}\rangle$. If we start with a known mixed state, then its
restoration after a projective measurement is a little more
involved;\cite{restore-known} however such a procedure still can be
easily analyzed using standard quantum mechanics and classical
probability theory.

    In this paper we consider a different, non-trivial situation:
we assume that an arbitrary initial state is {\it unknown to us},
and we still want to restore it after the measurement disturbance.
To make this idea more precise, we consider a contest between the
uncollapse {\it proponent} Plato, and an uncollapse {\it skeptic}
Socrates.  Socrates prepares a quantum system in any state he likes,
but it is unknown to Plato.   Socrates sends the state to Plato, who
makes some measurement on the system, verified by the {\it arbiter}
Aristotle.  Plato then tries to undo the measurement. If Plato
judges that the attempt succeeded, the system is returned to
Socrates, with the claim that it is in the original state. Socrates
is then allowed to try and find a contradiction in any way he likes,
with the whole process monitored by Aristotle.  If a contradiction
can be found, then he can claim to refute the uncollapse claim, but
in the absence of contradiction, the uncollapse claim stands. If
Socrates would like to try again to find a contradiction, or if
Plato judges that his undoing attempt was unsuccessful (and does not
return a state), then Socrates prepares a new (still unknown to
Plato) state, and the competition continues.   If Socrates cannot
find a contradiction after many rounds of the competition, then
Plato will win the contest, and will have successfully demonstrated
the uncollapsing of the quantum state.

    A slightly different but equivalent situation is when we know the
initial state, but our uncollapsing procedure must be independent of
the initial state (so we can pretend that it is unknown), and
therefore the uncollapsing should restore {\it any} initial state in
the same way. This formulation is most appropriate for a real
experiment demonstrating the uncollapsing. Finally, we may consider
the more general case where the measured system is entangled with
another system, and we wish to restore the initial state of the
compound system without any access to its second part.

The traditional statement of irreversibility of a quantum
measurement can be traced to the fact that it may be described as a
mathematical projection.  Projection is a many-to-one mapping in the
Hilbert space, and therefore the same post-measurement state
generally corresponds to (infinitely) many initial
states.\cite{exception}  It is therefore impossible to undo a
projective measurement.

However, the situation is different for a general \cite{Nielsen}
(POVM-type) measurement, which typically corresponds to a one-to-one
mapping $|\psi_{in}\rangle \rightarrow |\psi_m\rangle$ in the
Hilbert space of wavefunctions (in this paper we consider only
``ideal'' measurements which do not introduce extra decoherence). In
this case the post-measurement wavefunction $|\psi_m\rangle$ can
still be associated with the unique initial state
$|\psi_{in}\rangle$, and a well-defined inverse mapping exists
mathematically. This makes the uncollapsing possible in principle.
Since the inverse mapping is typically non-unitary, it cannot be
realized as an evolution with a suitable Hamiltonian. However, it
can be realized using another POVM-type measurement with a specific
(``lucky'') result.

\subsection{Formalism for wavefunctions}
\label{formalism-wf}

Let us first consider a pure initial state $|\psi_{in}\rangle$, and
postpone a generalization to mixed states until Sec. II\,C. In the
formalism of a general (ideal) quantum measurement \cite{Nielsen}
which transfers pure states into pure states, the measurement with
result $m$ is associated with the linear Kraus operator $M_m$, so
that the probability of result $m$ is
    \be P_m(|\psi_{in}\ra) =  || M_m |\psi_{in}\rangle \, ||^2 ,
    \label{prob-wf} \ee
where $|| \ldots ||$ denotes the norm of the state, and the (conditioned)
state after measurement is
    \be |\psi_m \rangle = \frac{M_m
    |\psi_{in}\rangle}{\sqrt{P_m(|\psi_{in}\ra)}} ,
    \label{state-wf} \ee
where the denominator makes $|\psi_m\rangle$ properly normalized.
(Very often people prefer to omit this denominator and work with
non-normalized states; this makes the mapping linear.) The operators
$E_m=M_m^\dagger M_m$ (called POVM elements \cite{Nielsen}) are
Hermitian and positive semidefinite by construction; these operators
must obey the completeness relation $\sum_m E_m ={\bf 1}$, which
ensures that the total probability of all measurement results is
unity. A measurement operator $M_m$ can always be written as \be
M_m=U_m\sqrt{E_m}, \label{mop} \ee where $U_m$ is a unitary operator
(an important special case is when $M_m=\sqrt{E_m}$; this
corresponds to the ``quantum Bayes theorem'' \cite{Gardiner}).

Now let us discuss wavefunction uncollapse in this general and
abstract context. The state disturbance rule (\ref{state-wf}) is
typically a nonunitary one-to-one map in the Hilbert space. To undo
the measurement with known result $m$, we have to realize a physical
process corresponding to the nonunitary inverse operator $M_m^{-1}$,
multiplied by an arbitrary constant (which is not important because
of the normalization). This can be accomplished with another
measurement, possibly together with unitary operations. As shown
below, we can realize measurement undoing if the second measurement
realizes a Krauss operator of the form
    \be L = C U_L E_m^{-1/2} V_L ,
    \label{L}  \ee
where $U_L$ and $V_L$ are any unitary operators, and
$C$ is an unspecified constant that will be discussed later. (The
operator $L$ also has a decomposition of the form $U\sqrt{E}$, but
with a different POVM element $E$.)
  The
uncollapse then consists of three steps: first, the unitary operator
$V_L^\dagger U_m^\dagger$ is applied to reverse the unitary part of
$M_m$ and prepare for the second measurement.  Next the measurement
operator $L$ is applied.  Finally the unitary operator $U_L^\dagger$
is applied to reverse the remaining unitary part of $L$. We can now
see the effect of the uncollapsing operation on the state
$|\psi_m\rangle$ by applying Eq.\ (\ref{state-wf}) and the unitaries
to find
    \be
|\psi_{f}\rangle = \frac{U_L^\dagger L V_L^\dagger
U_m^\dagger|\psi_m\rangle} {||  U_L^\dagger L V_L^\dagger
U_m^\dagger|\psi_m\rangle ||}
   = |\psi_{in} \rangle ,
    \ee
thus restoring the original state, because
$U_L^\dagger L V_L^\dagger U_m^\dagger M_m= C$, which is removed by
the
normalization. 
    (The phase of $C$ is not important, since it affects only the overall phase
of the wavefunction.)

However, in order for the operator $L$ to be physically realizable,
the operator $L^\dagger L$ must belong to another complete set of
POVM elements, and therefore all its eigenvalues must not exceed
unity (otherwise some states will be assigned probabilities that are
above unity; notice that the eigenvalues are non-negative
automatically). Since $L^\dagger L=|C|^2 V_L^\dagger E_m^{-1}V_L$,
its eigenvalues are directly related to the
eigenvalues $p_i^{(m)}$ of the operator $E_m$. Expressing
$E_m=\sum_i p_i^{(m)} |i\rangle \langle i|$, where the eigenvectors
$|i\rangle$ form an orthonormal basis, the eigenvectors of
$L^\dagger L$ are obviously $V_L^\dagger |i\rangle$, and the
corresponding eigenvalues are $|C|^2/p_i^{(m)}$. Since all these
eigenvalues must not exceed 1, we find the following inequality on
$|C|^2$,
    \be
|C|^2 \le \min_i p^{(m)}_i =  \min P_m,
    \label{bound-C} \ee
where $\min P_m$ is the probability of the result $m$, minimized
over all possible states $|\psi_{in}\ra$ in the Hilbert space.  The
equality of $\min P_m$ to $\min_i p^{(m)}_i$ follows from Eq.\
(\ref{prob-wf}).

  There is no guarantee that the uncollapse can be accomplished
deterministically, since we rely on a measurement with a specific
result, corresponding to the operator $L$. We can calculate the
uncollapse success probability $P_S$ from Eq.~(\ref{prob-wf}), with
$M_m \rightarrow L$ and $|\psi_{in}\ra \rightarrow V_L^\dagger
U_m^\dagger | \psi_m \ra$,
    \be
P_S = || LV_L^\dagger U_m^\dagger |\psi_m\rangle ||^2 = \left|\left| \frac{
C|\psi_{in}\rangle }{\sqrt{P_m(|\psi_{in}\ra)}} \right|\right|^2 =
\frac{|C|^2}{P_m(|\psi_{in}\ra)}.
    \label{P_S-1}    \ee
Now using the bound (\ref{bound-C}) for $|C|^2$, we find the bound
for the success probability of uncollapsing after the first
measurement with result $m$:
    \be
    P_S \leq \frac{\min P_m}{P_m (|\psi_{in}\rangle)} ,
    \label{P_S-wf}\ee
where the denominator is the probability of the result $m$ for a
given initial state, while the numerator is this probability minimized over all
possible initial states.

    The bound (\ref{P_S-wf}) is one of the most important results (notice
a similar result in Ref.\ [\onlinecite{Ueda-99}]) and deserves a
discussion.
    First, this bound is exact in the sense that it is achievable by
the optimal uncollapsing procedure. This is because the
uncollapsing operator with $|C|=\sqrt{\min P_m}$ is still a
physically allowed operator.
 As we will see later, the upper bound (\ref{P_S-wf}) is achievable in
real experimental setups. However, non-optimal uncollapsing
procedures, especially involving a sequence of measurements, can
lead to smaller success probabilities (an example of non-optimal
uncollapsing has been discussed in Ref. [\onlinecite{Ruskov-07}]).
An analysis of the procedures with an arbitrary sequence of
measurements and unitary operations is similar to the above: the
corresponding measurement and unitary operators should simply be
multiplied.

    Notice that the success probability (\ref{P_S-1}) and the
inequality ($\ref{P_S-wf}$) depends on the initial state, which is
unknown to the person performing the uncollapsing (Plato, see
description in the previous subsection). Therefore, the success
probability $P_S$ can be calculated by the man who knows what the
initial state is (Socrates), while Plato can only estimate $P_S$;
for example, he can calculate the worst-case scenario (the minimum
of $P_S$ over the accessible Hilbert space) or can calculate the
average of $P_S$ over all possible initial states (this procedure
will be discussed in the next subsection).

    Recalling the fact that it is not possible to undo a fully collapsed
state due to the nature of projective measurement, the uncollapsing
probability $P_S$ should decrease with increasing strength of the
first measurement. Qualitatively, a stronger measurement is one that
tends to a projection, as the uncertainty in the measurement
decreases.  Mathematically, this means that some eigenvalues of
$E_m$ become closer to 0.  As a consequence, $\min P_m$ becomes
smaller [see Eq.\ (\ref{bound-C})], therefore lowering the upper bound for
$P_S$.  For a projective measurement $P_S = \min P_m=0$, thus making
the state uncollapse impossible.

   It is interesting to discuss the case when the initial state $|\psi_{in}\rangle$
is known to belong to a certain subspace of the Hilbert space, and
we therefore wish to restore states only in this subspace. In this
case, the calculation of $\min P_m$ should be limited to this
subspace, which may increase the bound (\ref{P_S-wf}) for the
success probability $P_S$.  A trivial example of such a situation is
when the initial state $|\psi_{in}\rangle$ is known to Plato.  Here
it is not necessary to minimize $P_m$ over all possible initial
states in Eq.\ (\ref{P_S-wf}), because the set of possible states
consists of only one (known) state, thus allowing uncollapsing with
100\% probability. This is exactly the case discussed at the
beginning of Sec.\ \ref{PrDis}.

\subsection{Formalism for density matrices}
\label{formalism-dm}

 So far we have dealt only with pure states; however,
it is very simple to generalize the uncollapsing formalism to
include  density matrices. In this case the initial density matrix
$\rho_{in}$ is transformed by the first measurement into the state
\cite{Nielsen}
    \be
    \rho_m= \frac{M_m\rho_{in}M_m^\dagger}{P_m},
    \label{state-dm}\ee
where the probability $P_m$ of the measurement result $m$ is
    \be
    P_m(\rho_{in}) = \mbox{Tr}(M_m^\dagger M_m \rho_{in}).
    \label{prob-dm}\ee
Using the uncollapsing procedure previously discussed and using the
same measurement operator $L$ given by Eq.\ (\ref{L}), we find that
the uncollapsed state
    \be
   \rho_f = \frac{U_L^\dagger L V_L^\dagger U_m^\dagger \rho_m U_m V_L^\dagger
L^\dagger U_L}{\mbox{Tr}(L^\dagger L V_L U_m^\dagger \rho_m U_m
V_L^\dagger )} = \rho_{in}
    \label{rho-f}\ee
coincides with the initial state. The uncollapsing success
probability $P_S$ is equal to the denominator in Eq.\ (\ref{rho-f}),
and satisfies the relation
    \be
P_S=|C|^2/P_m (\rho_{in})
    \label{P_S-dm}\ee
 [as in Eq.\ (\ref{P_S-1})]. The constant $|C|^2$ is
still limited by the inequality (\ref{bound-C}), and therefore the
probability of success has the upper bound
    \be
P_S \le \frac{ \min P_m }{P_m(\rho_{in})},
    \label{gen-result} \ee
which is the same as the bound (\ref{P_S-wf}), except for the new
notation in the denominator, which reminds us of the possibly mixed
initial state. The minimization of $P_m$ in the numerator should now
be performed over the space of all possible initial mixed states;
however, the result obviously coincides with the minimization over
the pure states only.
    Similar to the case discussed in Sec.\ \ref{formalism-wf}, the
inequality (\ref{gen-result}) is the exact bound; it is achieved by
the optimal uncollapsing procedure, which maximizes $|C|$.

    If the initial state is pure, then the formalism of this
subsection is trivially equivalent to the formalism of Sec.\
\ref{formalism-wf}. It becomes more general in the case when the
``actual'' initial state is mixed; for example, this happens when
the initial state has been in contact with an unmonitored
environment or Socrates prepares a state by a blind random choice
from a set of pure states. A more interesting case for the result
(\ref{gen-result}) is when the measured system is entangled with
another system, which does not evolve by itself. Then the formalism
can be applied to the compound system; however, the measurement
probability $P_m$ depends only on the reduced initial density
matrix, traced over the entangled second part. Therefore, in the
entangled bipartite case the uncollapsing procedure restores the
state of the whole system, while the success probability $P_S$ is
given by Eq.\ (\ref{gen-result}) with $\rho_{in}$ being the reduced
density matrix.

    Another advantage of Eq.\ (\ref{gen-result}) in comparison with
Eq.\ (\ref{P_S-wf}) is the following. In the derivation of both
results the initial state is the ``actual'' initial state, which is
known to Socrates, but typically unknown to Plato. However, as we
will prove below, Eq.\ (\ref{gen-result}) can still be used by Plato
in a somewhat different sense: with $\rho_{in}$ being understood as
an {\it averaged} density matrix representing a distribution of
possible initial states.  In this case, Eq.\ (\ref{gen-result})
gives the uncollapsing probability averaged over this distribution.
For example, if Plato knew that Socrates' strategy is to prepare one
of 2 possible (nonorthogonal) states $|\psi_1\ra, |\psi_2\ra$, with
probabilities ${\cal P}$ and $1-{\cal P}$, then he could find the
average undoing probability in two ways.  The first method is that
he could simply average the undoing probabilities of the two states
(also taking into account the information acquired in the first
measurement, see below).  Alternatively, he could recall that the
random state preparation described above is equivalent to
considering the initial density matrix $\rho_{in} = {\cal P}
|\psi_1\ra\la \psi_1| + (1-{\cal P}) |\psi_2\ra\la \psi_2|$, and
then apply \cite{note} the result (\ref{gen-result}) to this density
matrix. In this way, in the absence of any information, Plato could
estimate his typical success rate by calculating (\ref{gen-result})
for a fully mixed state, invoking the principle of
indifference.\cite{Jaynes}

In the general case the above statement, that both ways of computing
the averaged undoing probability are equivalent, can be proven both
logically and explicitly. For the logical proof we notice that
Plato's judgment of successful undoing does not depend on whether or
not Socrates knows the randomly picked state; therefore, the average
probability of the cases judged to be successful should be the same
in both situations (whether or not Socrates knows what the state
is). Now let us also prove this statement explicitly, thus checking
that our formalism is self-consistent. Suppose the initial state is
prepared by Socrates by choosing randomly from a set of initial
states $\rho^{(k)}$ with probabilities ${\cal P}_k$ (the most
natural case is when initial states are pure,
$\rho^{(k)}=|\psi_k\rangle \langle \psi_k|$; however, this is not
necessary).  Then the bound for the average probability of
uncollapsing success $P_S^{(\rm av)}$ is the average of the bounds
(\ref{gen-result}):
    \be
P_S^{\rm (av)} \le \sum\nolimits_k  \frac{\min P_m}{P_m(\rho^{(k)})}
\, {\cal P}'_k .
\label{P-S-av}\ee
Notice, however, that ${\cal P}'_k$ is the posterior probability
distribution given the result $m$, which is different from ${\cal
P}_k$.
   We may now invoke the classical Bayes rule \cite{Feller,Jaynes}
    \be
{\cal P}(k|m) = \frac{{\cal P}(m|k) {\cal P}_k}{\sum_{\tilde k}
{\cal P}(m|\tilde k) {\cal P}_{\tilde k}}
    \ee
    to relate the posterior
${\cal P}'_k= {\cal P}(k|m)$ to the prior ${\cal P}_k$ and the
conditional probability ${\cal P}(m|k)=P_m(\rho^{(k)})$ to have
result $m$ given state $k$, so that
    \be
    {\cal P}'_k = \frac{P_m (\rho^{(k)}){\cal P}_k}
    {\sum_{\tilde k} P_m (\rho^{(\tilde k)}) {\cal P}_{\tilde k} }.
    \label{P'}\ee
Substituting Eq.\ (\ref{P'}) into Eq.\ (\ref{P-S-av}) and using $\sum_k
{\cal P}_k =1$ in the numerator, we obtain
    \be P_S^{({\rm av})} \le \frac{\min P_m}{\sum_k P_m(\rho^{(k)}) {\cal
    P}_k}= \frac{\min P_m}{P_m(\rho^{({\rm av})})},
    \ee
where $ \rho^{(\rm av)}= \sum_k \rho^{(k)} {\cal P}_k$ is the
averaged initial state. This ends the proof that Eq.\
(\ref{gen-result}) can be used for an unknown initial state, with
$\rho_{in}$ being understood as the average of all possible initial
states.

\subsection{Uncollapsing probability, information, and an irreversibility measure}
\label{sect-informat}

    We defined the success probability $P_S$ as a
probability to uncollapse the post-measurement state $\rho_m$. We now
wish to start counting the overall success probability $\tilde P_S$
from the time before the first measurement, so that $\tilde P_S$ is the probability
of the pair of events: measurement with result $m$ and then
successful uncollapsing. Using Eqs.\ (\ref{P_S-dm}) and
(\ref{gen-result}) we easily find the relation
    \be
    \tilde P_S = P_m(\rho_{in}) P_S = |C|^2
    \label{tildeP}
    \ee
and the upper bound
    \be
    \tilde P_S \leq \min P_m .
    \label{tilde-P_S}\ee
Notice that $\tilde P_S$ is independent of the initial state. While
this property seems to be somewhat surprising, we will see later why
it is rather obvious.

    If we now wish to consider all possible results of the first
measurement, and perform different uncollapsing procedures for each
measurement result, then the total probability of uncollapsing
$\tilde P_S^{\rm total}$ is bounded as
   \be
    \tilde P_S^{\rm total} \leq \sum\nolimits_m \min P_m ,
    \label{tilde-P_S-total}\ee
and is also independent of the initial state.

The bounds (\ref{tilde-P_S}) and (\ref{tilde-P_S-total}) are exact
and reachable by optimal uncollapsing procedures. The bound
(\ref{tilde-P_S-total}) indicates that $1-\sum_m \min P_m$ can be
used as a {\it measure of irreversibility} (collapse strength) due
to the measurement operation.

    Now let us discuss the relationship between the uncollapsing procedure
and our knowledge of the initial state. While uncollapsing is
possible even if we know nothing about the initial state of the
system, at first glance it seems like we gain some knowledge about
the initial state in the process. This leads to the following
interesting paradox, initially considered by Royer.\cite{Royer}  By
doing both a measurement and unmeasurement, one can seemingly learn
something about the initial state without disturbing it.  Then by
repeating the measurement + unmeasurement process many times, even
though the probability of such an event rapidly decreases to zero,
the successful event would lead to essentially perfect knowledge of
the initial state, leaving the state itself perfectly intact!  One
could then violate a host of known results, such as the no-cloning
theorem.  The resolution of the paradox lies in the fact that the
pair of measurement and unmeasurement actually brings exactly {\it
zero information}. Uncollapsing the state can only occur when the
information in the second measurement exactly contradicts the
information gained in the first measurement, thus nullifying it.
This can happen in weak quantum measurements because there is
uncertainty about the system in the measurement result.  It is to
the extent that this ambiguity exists that it is possible to undo
the weak measurement. Let us examine this in more detail.

    We learn something about a pre-measurement state when the measurement
result depends on the state. The measurement with result $m$ brings
some information about $\rho_{in}$ because the probability
$P_m(\rho_{in})$ depends on the initial state $\rho_{in}$. The
ability to successfully uncollapse the state {\it also} brings some
information about the initial state because the uncollapsing
probability $P_S=|C|^2/P_m(\rho_{in})$ also depends on $\rho_{in}$.
However, the collapse-uncollapse probability ${\tilde P}_S$ of
observing {\it both} the result $m$ followed by a successful
uncollapse is independent of $\rho_{in}$ -- see Eq.~(\ref{tildeP}).
Therefore, the combined effect of partial collapse
and uncollapse brings no information about the initial state.

More quantitatively, we can use the same framework as at the end of
Sec.\ \ref{formalism-dm} in order to track the information gain
during the procedure.  Suppose Plato assigns an
initial distribution ${\cal P}_k$ of possible initial states as a
statistical prior, to be updated as more information comes in.  The
measurement with result $m$ brings in this information, so Plato
updates his prior to the posterior distribution ${\cal P}'_k$ [see
Eq.\ (\ref{P'})]. Calculating in a similar way the distribution
${\cal P}''_k$ after the pair of the measurement and unmeasurement
results, we find
    \be
      {\cal P}''_k = \frac{P_S (\rho^{(k)}){\cal P}'_k}
    {\sum_{\tilde k} P_S (\rho^{(\tilde k)}) {\cal P}'_{\tilde k} }
     =  \frac{\tilde P_S (\rho^{(k)}) {\cal P}_k}
    {\sum_{\tilde k} \tilde P_S (\rho^{(k)}) {\cal P}_{\tilde k} },
    \label{p''}
    \ee
where $P_S(\rho)$ and $\tilde P_S(\rho)$  denote respectively the
probabilities of uncollapsing (\ref{gen-result}) and a combined
collapse-uncollapse pair (\ref{tildeP}) for the initial state
$\rho$.  However, as we have already stressed, $\tilde P_S(\rho)$ is
independent of the initial state $\rho$, and therefore cancels out
of the expression (\ref{p''}).  This fact (and the normalization of
the prior $\{P_k\}$) restores the initial prior distribution, ${\cal
P}''_k ={\cal P}_k$, and therefore Plato has learned nothing, thus
avoiding the paradox. Reversing the logic, in order to avoid the
paradox, $\tilde P_S$ must be independent of the initial state, as
found in Eqs.~(\ref{tildeP}) and (\ref{tilde-P_S}).

\section{
    Double-quantum-dot charge qubit}


Consider Fig.\ \ref{ok} illustrating a double-quantum-dot (DQD)
qubit, measured continuously by a symmetric quantum point contact
(QPC). This setup has been extensively studied in earlier papers,
both theoretically\cite{DDth} and experimentally.\cite{DDexp} The
measurement is characterized by the average currents $I_1$ and $I_2$
corresponding to the qubit state $|1\rangle$ and $|2\rangle$ (the
double-dot electron being in one dot, or the other), and by the shot
noise spectral density $S_I$. \cite{S-normalization} We treat the
additive detector shot noise as a Gaussian, white, stochastic
process, and assume the detector is in the weakly responding regime,
$|\Delta I| \ll I_0$, where $\Delta I=I_1-I_2$ and
$I_0=(I_1+I_2)/2$, with QPC voltage much larger than all other
energy scales,
so that the measurement process can be described by the
quantum Bayesian formalism.\cite{Kor-99}

\subsection{Measurement dynamics for a non-evolving
qubit}\label{charge-dynamics}

We begin for simplicity with the assumption that there is no qubit
Hamiltonian evolution,
 so that the qubit state evolves due to the measurement only
(this can also be effectively done using ``kicked'' quantum
nondemolition (QND) measurements \cite{Jordan}). As was shown in
Ref.~\onlinecite{Kor-99}, at low temperature the QPC
is an ideal quantum detector (which does not decohere the measured
qubit), so that the evolution of the qubit density matrix $\rho$ due
to continuous measurement preserves the ``murity''\cite{ourprb}
${\cal M}$ while the diagonal matrix elements evolve according to
the classical Bayes rule.\cite{Feller} We define the electrical
current through the QPC averaged in a time $t$ as
$\bar{I}(t)=[\int_0^t I(t')\, dt']/t$, and the quantum Bayesian
equations read
    \bea
\rho_{11}(t) &=& \frac{\rho_{11}(0) P_1({\bar I})}
{\rho_{11}(0) P_1({\bar I}) + \rho_{22}(0) P_2({\bar I})}, \qquad
    \label{rho11simple} \\
\rho_{22}(t) &=& \frac{\rho_{22}(0) P_2({\bar I})} {\rho_{11}(0)
P_1({\bar I}) + \rho_{22}(0) P_2({\bar I})}, \qquad
    \label{rho22simple}\\
{\cal M} &=& \rho_{12}/\sqrt{\rho_{11}\rho_{22}} = {\rm const},
    \label{murity}
    \eea
where the conditional (Gaussian) probability densities of a current
$\bar{I}$ realization, given that the qubit is in $\vert 1 \ra,
\vert 2\ra$ are
    \be
  P_{1,2}({\bar I})= \sqrt{t/\pi S_I}\, \exp [-({\bar{I}-I_{1,2}})^2 t/S_I].
  \label{P1,2}
    \ee
Equations (\ref{rho11simple}) and (\ref{rho22simple}) may be
simplified by noting
     \begin{equation}
\frac{\rho_{11}(t)}{\rho_{22}(t)} = \frac{\rho_{11}(0)}{\rho_{22}(0)}\, e^{2r(t)},
    \label{Bayes-DD}
    \end{equation}
where we define the dimensionless {\it measurement result} as
    \be
r(t)= \frac{t\Delta I}{S_I} \, [\bar{I}(t)-I_0]  = \frac{\Delta
I}{S_I} \int_0^t [I(t')-I_0]\, dt'.
    \label{result} \ee
    Notice that $r(t)$ is closely related to the total charge passed
through the QPC, and therefore $r(t)$ accumulates in time.
    For times much longer than the ``measurement time'' $T_M= 2 S_I/(\Delta I)^2$
(the time scale required to obtain a signal-to-noise ratio of 1),
the average current $\bar{I}$ tends to either $I_1$ or $I_2$ because
the probability density $P({\bar I})$ of a particular $\bar{I}$ is
    \be
    P({\bar I})=\sum_{i=1,2}
 \rho_{ii}(0) P_i({\bar I}).
\label{output}
    \ee
Therefore $r(t)$ tends to $\pm \infty$, continuously collapsing the state to
either $|1\rangle$ (for $r\rightarrow \infty$) or $|2\rangle$ (for
$r\rightarrow -\infty$).   Importantly, for the special case when the initial
state is pure, the state remains pure during the entire process.

This set of DQD measurement dynamics can be seen to be related to the
general measurement formalism \cite{Nielsen} described in the
previous section in the following way.\cite{ourprb} For a fixed time
$t$ the measurement result $m$ can be associated with the averaged
current $\bar{I}$ (or, equivalently, with the dimensionless quantity
$r$). The Kraus operator $M_m$ is then diagonal in the measurement
basis $|1\rangle$ and $|2\rangle$ and has matrix elements
$\sqrt{P_{1,2}({\bar I})}$. Notice that $P_m$ now describes the
probability density of the result $\bar I$ instead of probability,
because the measurement result becomes a continuous variable.

\subsection{Uncollapsing for the charge qubit}
\label{charge-uncollapse}

\label{uncollapsing-charge}

In order to describe how to uncollapse the charge qubit state, we
note that if $r(t)=0$ at some moment $t$, then the qubit state
becomes exactly the same as it was initially, $\rho (t)=\rho(0)$, as
follows from Eqs.\ (\ref{Bayes-DD}) and (\ref{murity}). This of
course must be the case if $t=0$, {\it i.e.}\ before the measurement
began, but is equally valid for some later time.  To see why this is
so from the informational point of view, we note that in the absence
of noise, the measurement result from states $\vert 1\ra, |2\ra$
would simply be $r_{1,2}(t) =\pm t/T_M$. With the noise present, the
measurement outcome $r(t)=0$ splits in half the difference between
states $\vert 1 \ra$ and $\vert 2\ra$. Such an outcome corresponds
to an equal statistical likelihood of the states $\vert 1\rangle$
and $\vert 2\rangle$, and therefore provides no information about
the state of the qubit.

\begin{figure}[tb]
  \centering
 \includegraphics[width=8cm]{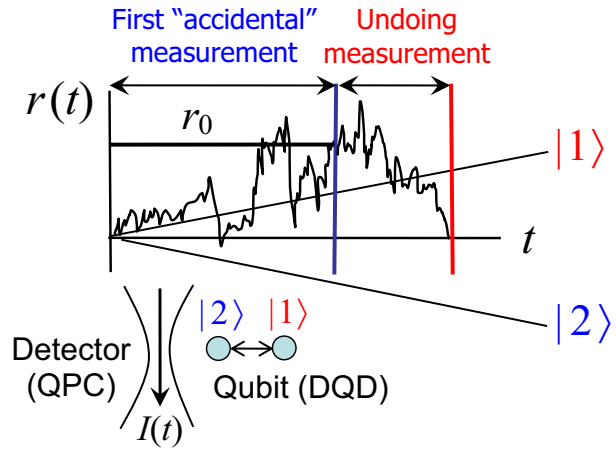}
  \caption{(Color online).
  Illustration
  of the uncollapsing procedure for the charge qubit.  The slanted
  lines indicate the deterministic output of the detector in the absence
  of noise, if the qubit is in state $\vert 1\ra$ or $\vert 2 \ra$.
  The initial measurement yields the result $r_0$. The detector is
  again turned on, hoping that at some future time the measurement result
  $r(t)=r_0+r_u(t)$ crosses the origin, at which time the detector is turned off,
  successfully erasing the information obtained in the first measurement, and
  restoring the initial qubit state. }
  \label{ok}
\end{figure}

Suppose the outcome of a measurement is $r_0$, partially collapsing
the qubit state toward either state $\vert 1\ra$ (if $r_0 >0$), or
state $\vert 2\ra$ (if $r_0 <0$).  The previous ``no information"
observation suggests the following strategy for uncollapsing:
continue measuring, with the hope that after some time $t$ the
stochastic result of the second measurement $r_u(t)$ becomes equal
to $-r_0$, so the total result $r(t)=r_0+r_u(t)$ is zero, and
therefore the initial qubit state is fully restored. If this
happens, the measuring device is immediately switched off and the
uncollapsing procedure is successful (Fig.\ \ref{ok}).  However
$r(t)$ may never cross the origin, and then the uncollapsing attempt
fails.

This strategy requires the observation of a particular measurement
result that may never materialize.  The strategy shifts the
randomness to the amount of time that needs to elapse in order to
find the desired measurement result.  Of course, in a given
realization the measurement result could take on the desired value
multiple times, so we will take as our strategy to turn off the
detector the first time the measurement result takes on $r=0$. In
the classical stochastic physics this is known as a first passage
process, \cite{redner} the theory of which is well developed and
will be used below.

In order to analyze the uncollapsing strategy performance, in
particular to find the success probability $P_S$, it is important to
notice that the off-diagonal elements of the qubit density matrix
$\rho$ do not come into play when we consider the detector output
$I(t)$ (this is true only in the case of zero or QND-eliminated
qubit Hamiltonian; we will shortly generalize to the finite qubit
Hamiltonian case).  As a result, the quantum problem can be exactly
reduced to a classical problem by substituting $\rho_{11}(t)$ and
$\rho_{22}(t)$ with classical probabilities, evolving in the course
of measurement according to the classical Bayes rule, while
evolution of the off-diagonal elements $\rho_{12}=\rho_{21}^*$ can
be found automatically from the murity conservation law
(\ref{murity}). We therefore model the qubit by a classical bit with
probability $p_1=\rho_{11}(0)$ of being prepared in state ``1'' and
probability $p_2=\rho_{22}(0)$ of being in state ``2''.
If the bit is in state ``1'', the dimensionless measurement result $r(t)$ evolves as a
random walk with diffusion coefficient $D=(\Delta I)^2/4S_I = 1/2T_M$ and
drift velocity $v_1=(\Delta I)^2/2S_I = 1/T_M$ [see Eqs.\ (\ref{P1,2}) and
(\ref{result})]. For the bit state ``2'' the random walk of $r(t)$
has the same diffusion coefficient but the opposite drift velocity
$v_2=-(\Delta I)^2/2S_I$. We are given the fact that the first part
of the measurement had the result $r_0$ (i.e.\ we select only such
realizations). We need to analyze the stochastic behavior of the
total measurement result $r(t)$ during the second part of
measurement, with most attention to the crossing of the zero line
$r(t)=0$ [for convenience we shift $t=0$ to the beginning of the
second measurement, so that $r(0)=r_0$].

    Let us find the probability $P_S$ of such a crossing. We will
first obtain it in a simple way, and then reproduce the result in a
more complicated way, which will also allow us to analyze the
statistics of the waiting time. For definiteness take $r_0>0$ (this
will be extended later). Then the result $r(t)$ will necessarily
cross 0 if the bit is actually in the state ``2'', because in this
case $r(t=\infty )=-\infty$ while $r(t=0)=r_0>0$. The probability of
being in the state ``2'' is $\tilde{p}_2=p_2
e^{-r_0}/(p_1e^{r_0}+p_2 e^{-{r_0}})$ from (\ref{Bayes-DD}), which
differs from $p_2$ because of the Bayesian update. If the bit is in
state ``1'' (this happens with probability $\tilde p_1=1-\tilde
p_2$), then $r(t=\infty )=+\infty$ and the crossing of 0 may never
happen; however, it is still possible with some probability $P_C$,
which depends on $r_0$, and also on $D$ and $v_1$. To find $P_C$,
let us consider an infinitesimal time step $dt$ and model the
diffusion by discrete jumps in $r$ of magnitude $\Delta r =\pm
\sqrt{2D\, dt}$. After a step $dt$, the coordinate will then shift
to one of two positions, $r=r_{\pm}$, where $r_\pm = r_0 + v_1 dt
\pm \sqrt{2D\, dt}$.  Each of these new coordinates will have its
own probability of eventually crossing the origin, $P_C(r_\pm)$.
Because the diffusive dynamics is generated by choosing either $r_+$
or $r_-$ with equal weighting, it follows in the limit ${dt
\rightarrow 0}$ that
    \be
    P_C(r_0)=  \sum _\pm \frac{1}{2}\, P_C(r_\pm).
    \ee
Expanding $P_C(r_\pm)$ in this relation in a Taylor series, we
find from the linear in $dt$ term that
    \be
    D\, \partial_{r_0}^2 P_C= -v_1\, \partial_{r_0} P_C,
     \ee
where $\partial_{r_0}$ and $\partial_{r_0}^2$ denote the first and
second derivatives with respect to $r_0$.
     Taking into account that $P_C=1$ for $r_0=0$ and  $P_C=0$ for
$r_0=\infty$, the above differential equation may be easily solved
to find $P_C=\exp(-v_1 r_0/D)=\exp(-2r_0)$. Now collecting the
probabilities of the zero line crossing for both bit states, we find
    \be
P_S=\tilde p_1 P_C+\tilde p_2=e^{-r_0}/(p_1 e^{r_0}+p_2 e^{-r_0}).
    \ee
The derivation for $r_0<0$ is similar and leads to the extra factor
$e^{2r_0}$, so that the crossing probability in both cases can be
written as $P_S=e^{-|r_0|}/(p_1 e^{r_0}+p_2 e^{-r_0})$.

    Thus, using the trick of reducing the quantum dynamics to the
classical problem, we have found the probability of successful
uncollapsing for a DQD qubit with no Hamiltonian evolution:
    \be
    P_S=\frac{e^{-|r_0|}}{e^{r_0} \rho_{in,11} + e^{-r_0}
    \rho_{in,22}},
    \label{s} \ee
where $\rho_{in}$ characterizes the qubit state before the first
measurement.
    We will discuss this result in the next subsection. Before that
let us rederive it in a different way, using the power of the
standard methods of first passage theory.\cite{redner} This method
has recently been used to investigate entanglement dynamics of
jointly measured qubits.\cite{williams}

It is convenient to scale time in units of the measurement time,
$\tau \equiv t/T_M$; then the probability distributions (\ref{P1,2})
take the simple form
    \be P_{1,2}(r, \tau) = \sqrt{\frac{1}{2 \pi \tau}}
\exp\left( - \frac{(r \mp \tau)^2}{2 \tau} \right).
    \ee
    These are
the solutions of two different classical random walks with
dimensionless drift velocity $\tilde v_{1,2} = \pm 1$ and
dimensionless diffusion coefficient $\tilde D = 1/2$ described by
the Fokker-Planck equations,
    \be
\partial_\tau P_i(r,\tau) = - \tilde v_i \partial_r P_i + \tilde D \partial_r^2 P_i.
    \label{fp} \ee
    In order to solve the first passage problem, we solve
first for the Green functions $G(r, \tau )$ of the above equations
starting from the initial condition $r=r_0$. The solutions from the
different drift velocities will be weighted with probabilities
${\tilde p}_{1,2}$. These Green function equations are supplemented
with an absorbing boundary condition at the origin ($r=0$),
    \be
G_i(r=0, \tau ) = 0,
    \label{bc} \ee
in order to account for the statistics of events that cross this
point at least one time. Let as again start with assuming $r_0>0$
and consider the other case later. The solution of Eq.~(\ref{fp})
subject to the condition (\ref{bc}) is most easily found by
guessing:
    \bea
G_i(r, \tau) &=& \frac{1}{\sqrt{4 \pi \tilde D \tau}} \left(
\exp\left[
-\frac{( r - r_0 - \tilde v_i \tau)^2}{4 \tilde D \tau}\right] \right.
    \\
&-& \left.\exp\left[-\tilde v_i r_0 /\tilde D\right] \exp\left[-\frac{( r +
r_0 - \tilde v_i \tau)^2}{4 \tilde D \tau} \right] \right).
    \nonumber \label{ans} \eea
In the form written above, it is obvious that the solution obeys the
equation of motion (\ref{fp}) and has the correct initial condition
$r_0$ at $\tau=0$ (because the absorbing boundary condition only
permits $r \ge 0$ solutions).  Further inspection of the solution is
facilitated by factoring out the free Green function, $G_{\rm
free}(r, \tau ) = \exp[ - (r-r_0-\tilde v_i \tau)^2/(4 \tilde D
\tau)]/\sqrt{4 \pi \tilde D \tau}$ to write the solution as
    \be
G_i(r, \tau) = G_{\rm free}(r, \tau ) \left[1 - \exp\left(-
\frac{r\, r_0}{\tilde D \tau}\right)\right].
    \ee
    One can now explicitly see
that the absorbing boundary condition (\ref{bc}) is satisfied, and
the solution is completely positive (as it must be to represent a
probability density).

To calculate the first passage time distribution, we first note that the
total survival probability that the random walker will be in the
interval $r \in (0, \infty)$ at time $\tau$ is given by $P_{\rm
sur}(\tau) = \int_0^\infty dr G(r, \tau)$.  However, the only place
for the particle to be lost from the system is at the origin.
Therefore, the first passage time distribution $P_{\rm fpt}^{(i)}$ is given by
    \be
P_{\rm fpt}^{(i)} = -\partial_\tau  P_{\rm sur} = - \int_0^\infty dr
\partial_t G_i(r, \tau).
    \label{fptdist} \ee
    The next step is to note that the Fokker-Planck equation (\ref{fp})
may be rewritten as a continuity equation, $\partial_\tau G_i +
\partial_r J_i =0$. This simply means that locally, probability is
conserved.  The probability current in the continuity equation is
$J_i = -\tilde D
\partial_r G_i + \tilde v_i G_i$ from (\ref{fp}).   Substituting this into
(\ref{fptdist}) we find the general result \be P^{(i)}_{\rm
fpt}(\tau) = \int_0^\infty dr \partial_r J_i = J_i(\infty) - J_i (0)
= - J_i(0), \ee because the probability current at infinity
vanishes.  Applied to our problem, we find
    \be P^{(i)}_{\rm
fpt}(\tau) = \frac{r_0}{\sqrt{4 \pi \tilde D \tau^3} } \exp \left[ -
(r_0 + \tilde v_i \tau)^2/(4 \tilde D \tau) \right].
    \label{fptd} \ee
    The probability $P_C$
that the point $r=0$ is ever crossed is found by integrating
(\ref{fptd}) over all positive time to obtain
    \be
P_{C}= \begin{cases} \exp (-\tilde v_1 r_0/\tilde D)=\exp (-2r_0), & i=1, \\
 1, & i=2. \end{cases}
\label{cp}
\ee
 This result may be understood intuitively because if the state is in $i=2$ then
the drift $\tilde{v}_2=-1$ causes $r(t)$ to evolve from $r_0$ to
$-\infty$ and therefore must cross $0$ at some time, while if the
system is in state $i=1$ then the drift $\tilde{v}_1= 1$ causes
$r(t)$ to evolve from $r_0$ to $+\infty$. Therefore, in order to
cross $r=0$, the noise term must fight against the drift, causing a
successful crossing only occasionally.

In order to obtain the normalized first passage distribution
(conditioned on crossing), we divide (\ref{fptd}) by the
probabilities (\ref{cp}) to obtain
    \be
    P^{(i)}_{\rm fpt}(\tau |C)= \frac{r_0}{\sqrt{4 \pi \tilde D \tau^3} } \exp \left[ - (r_0 -
|\tilde v_i| \tau)^2/(4 \tilde D \tau) \right].
    \label{condfpt} \ee
    The mean first passage time may also be calculated from
(\ref{condfpt}) to obtain $\tau_{c,i}=r_0/|\tilde v_i|=r_0$.

Obtaining analogous results for $r_0 < 0$ is straightforward because
the Green function for the Fokker-Planck equation (\ref{fp}) is
invariant under the transformation $\{r \rightarrow -r, \, r_0
\rightarrow -r_0, \, \tilde v_i \rightarrow -\tilde v_i\ ({\rm or}\
1 \leftrightarrow 2)\}$ which is also reflection symmetry about the
origin.  Therefore results (\ref{fptd},\ref{cp},\ref{condfpt}) can
be extended using this symmetry.  Combining results, we can now
calculate the total uncollapsing probability, $P_S = {\tilde p}_1
P_{C,1} + {\tilde p}_2 P_{C,2}$ to obtain the result (\ref{s}) in
this new, more powerful way.

\subsection{Discussion}\label{charge-discussion}

We now discuss the physical meaning of the result (\ref{s}). When
the first measurement result indicates a particular qubit state with
good confidence ($|r_0|\gg 1$), the probability of success $P_S$
given by Eq.\ (\ref{s}) becomes very small, eventually becoming
$P_S=0$ for a projective measurement, realized for $r_0=\pm \infty$.
This recovers the traditional statement of irreversibility. In the
other limit of $r_0=0$, the success probability is unity because no
time needs to elapse - the state is already undisturbed. We stress
that the possibility of uncollapsing as well as our formalism
requires a quantum-limited detector, i.e. one that introduces no
additional dephasing to the system. For such a detector measuring a
pure state, the state remains pure throughout the partial collapse,
and the uncollapse. We also note that if the qubit is entangled with
other qubits, the uncollapsing restores the state of the whole
system.

Let us compare the general upper bound (\ref{gen-result}) for
the success probability $P_S$ with the result (\ref{s}). Substituting
the probabilities in the bound (\ref{gen-result}) with probability
densities, we find
    \be
    P_S \leq \frac{\min \{ P_1 (\bar{I}),P_2(\bar{I})\}}
    {P_1(\bar{I})\rho_{in,11}+P_2(\bar{I})\rho_{in,22}},
    \label{undo-general} \ee
where $\bar{I}$ corresponds to the measurement result $r_0$. The
bound coincides with Eq.\ (\ref{s}) because
$P_1(\bar{I})/P_2(\bar{I})=e^{2r_0}$. This means that the ``wait and
stop'' strategy analyzed above is optimal in the sense that it
reaches the upper bound.

It is also instructive to not specify the result of the first
measurement, but to find the total probability $\tilde P_S^{\rm
total}$ [see Eq.\ (\ref{tilde-P_S-total})] that the initial qubit
state can be restored after a measurement for time $t$. This is
given by averaging $P_S$ in Eq.\ (\ref{s}) over the results $r_0$
with the corresponding weights (\ref{output}). This averaging is
technically easier using the form of Eq.\ (\ref{undo-general}) and
gives
    \be
\tilde P_S^{\rm total} =  \int d{\bar I} \min \{P_1 ({\bar
I}),P_2({\bar I})\} = 1 -
\mbox{erf}\left(\sqrt{\frac{t}{2T_M}}\right),
    \label{tilde-P-S-tot-DQD}\ee
    which depends only
on the ``strength'' $t/T_M$ of the first measurement, but not on the
initial state, as expected from the discussion in Sec.\
\ref{sect-informat}. Notice that the result
(\ref{tilde-P-S-tot-DQD}) reaches the upper bound
(\ref{tilde-P_S-total}) because (\ref{s}) reaches the upper bound
(\ref{gen-result}).

\subsection{Statistical features of the time to
uncollapse}\label{charge-statistics}

In addition to the probability of success, the complete solution of
the first-passage problem given above now allows us to specify
further information about the uncollapsing process. In particular,
an important question for an experimental implementation of this
idea is how long it is necessary to wait.

    Since the distribution of the first passage time (\ref{condfpt})
does not depend on the bit state, it directly gives the distribution
of the waiting time to uncollapse any qubit state. Therefore,
rescaling back the time axis in Eq.\ (\ref{condfpt}), we find the
waiting time distribution is
    \be
    P_{\rm wait}(t)=  \frac{|r_0|}{\sqrt{2 \pi t^3/ T_M} } \exp \left[\frac{- (|r_0|-
t /T_M)^2}{2 t/T_M}\right].
    \label{undo_dist} \ee
This distribution is normalized, since we consider only successful
attempts of uncollapsing. The fact that the distribution is
independent of the initial qubit state is not surprising, since
otherwise a successful uncollapsing instance would give us an information about
the qubit state (see discussion in Sec.\ \ref{sect-informat}).

    Using the distribution (\ref{undo_dist}), we can find the mean waiting time to
uncollapse
    \be T_{\rm wait} = T_M  \, \vert r_0\vert,
    \label{wait} \ee
the standard deviation
    \be \Delta T_{\rm wait} = T_M \sqrt{|r_0|},
        \ee
and the most likely waiting time (which maximizes $P_{\rm wait}$)
    \be
    T_l = T_M \left(\sqrt{r_0^2 + 9/4} - 3/2\right) .
    \ee

\begin{figure}[tb]
  \centering
\includegraphics[width=8cm]{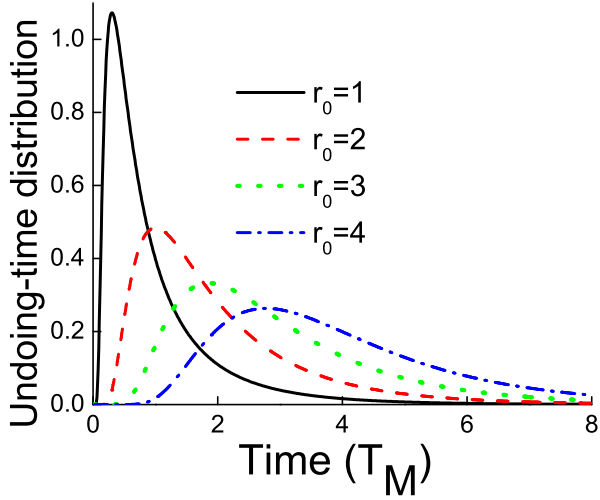}
  \caption{(Color online).  Probability distribution of the time
  required to undo the measurement. Different plots are for different
  values of $r_0$.}
  \label{fig-waiting}
\end{figure}

    The distribution (\ref{undo_dist}) of the waiting times is plotted in
Fig.~\ref{fig-waiting} for several values of $r_0$. Note that it has
a long tail, which makes the average value $T_{\rm wait}$ to be
longer than the most likely value $T_l$.

\subsection{Evolving charge qubit}\label{charge-evolve}

Let us now extend the first example of a single charge qubit
measured by QPC, by including internal evolution of the qubit via a
qubit Hamiltonian,
    \be
    H_{QB} = -(\varepsilon/2) \sigma_z + H \sigma_x,
    \ee
    where $\varepsilon$ is the energy asymmetry between
the quantum dot levels, and $H$ is the tunnel coupling between the
dots.  In this case Eqs.\ (\ref{rho11simple})--(\ref{murity}) are no
longer valid and should be replaced by the Bayesian equations
\cite{Kor-99} (in Stratonovich form\cite{Oksendal})
    \bea
    &&  \hspace{-8mm}
\dot{\rho}_{11}= -\dot{\rho}_{22}= -2\,H\,\mbox{Im}\,\rho_{12}
         +\rho_{11}\rho_{22}\, \frac{2\Delta I }{S_I}\, [I(t)-I_0],
        \label{Bayes1}\\
&& \hspace{-8mm} {\dot\rho}_{12}=  \imat \varepsilon\rho_{12}+
        \imat  H  (\rho_{11}-\rho_{22})
       \nonumber \\
&& {}
  -( \rho_{11}-  \rho_{22})  \frac{\Delta I}{S_I} \, [I(t)-I_0]\,
\rho_{12} ,
        \label{Bayes2}
    \eea
where $I(t)$ is the QPC current,
    \be
    I(t) = \rho_{11}(t) I_1 +\rho_{22}(t) I_2 +\xi(t),
    \ee
containing white noise $\xi(t)$ with spectral density $S_I$, and we
use $\hbar =1$.
    These evolution equations are nonlinear and not very simple to deal
with. To discuss the undoing of a continuous measurement, it is more
convenient to use a non-normalized density matrix $\sigma$, which
has an advantage of dealing with linear equations.

    We rewrite Eqs.\
(\ref{Bayes1})--(\ref{Bayes2}) in the form
\begin{eqnarray}
&&     \rho = \sigma /\mbox{Tr}\sigma ,
    \label{B-l-rho} \\
 && \dot{\sigma}_{11} = -2 H
\, \mbox{Im}\sigma_{12} -\sigma_{11} \frac{1}{S_I} [I(t)-I_1]^2  ,
\label{B-l-1}\\
&& \dot{\sigma}_{22} = 2 H \, \mbox{Im}\sigma_{12} -\sigma_{22} \frac{1}{S_I}
[I(t)-I_2]^2  ,
\label{B-l-2}\\
&&
 \dot{\sigma}_{12} = \imat \varepsilon\sigma_{12}
+i H (\sigma_{11}-\sigma_{22})
    \nonumber \\
&& \hspace{1cm} -\sigma_{12} \left\{ \frac{[I(t)-I_0]^2}{S_I} +\frac{(\Delta
I)^2}{4S_I}\right\} ,
    \label{B-l-3}
    \end{eqnarray}
so that $\sigma (0)=\rho (0)$, while the ratio $\sigma(t)/\rho(t)$
decreases with time and is equal to the normalized probability
density of the corresponding realization of the detector output
$I(t')$, $0\leq t'\leq t$.\cite{Nazarov-discuss}
      In the language of general quantum measurement this
formulation corresponds to omitting the denominator in Eq.\
(\ref{state-dm}). Notice that we still consider an ideal detector,
so an initially pure state remains pure,
$|\sigma_{12}|^2=\sigma_{11}\sigma_{22}$.\cite{Nazarov-discuss}

A casual inspection of Eqs.\ (\ref{B-l-1})-(\ref{B-l-3}) shows that
they are seemingly not well-defined because the terms
$[I(t)-I_{0,1,2}]^2$ contain the term $\xi(t)^2 = \infty$ [from the
relation $\la \xi(t) \xi(0)\ra = (S_I/2) \delta(t)$]. This
divergence is artificial because there will always be a small
correlation time $T$ of the noise and/or a finite detector bandwidth
$B$ (corresponding to $T=1/4B$), so there will be a large but finite
constant ${\cal C}= \langle \xi(t)^2\rangle = S_I/4T$ contained in
terms of the form $[I(t)-I_{0,1,2}]^2$. It is easy to see that Eqs.\
(\ref{B-l-1})-(\ref{B-l-3}) do not change if we subtract the same
constant from these terms $[I(t)-I_{0,1,2}]^2 \rightarrow
[I(t)-I_{0,1,2}]^2- {\cal C}$. This can be shown by considering
another unnormalized density matrix $\eta = \sigma \exp(t/T)$.
Writing the linear Bayesian equations (\ref{B-l-1})-(\ref{B-l-3}) in
the form ${\dot \sigma}_{ij} = f_{ij}[\sigma]$, the equations
transform to ${\dot \eta}_{ij} = f_{ij}[\eta \exp(-t/T)]\, \exp(t/T)
+ \eta_{ij}/T$ under the change of variables.  The unnormalized
Bayesian equations are linear in the density matrix elements
$\sigma_{ij}$, so the exponential factors cancel out. The new
equations are thus the same as the old ones with a constant ${\cal
C} = S_I/4 T$ subtracted from the $[I(t)-I_{0,1,2}]^2$ terms. The
unspecified constant $T$ in the density matrix transformation may be
chosen to be the short correlation time $T$ discussed above, thus
canceling the large term and making Eqs.\
(\ref{B-l-1})-(\ref{B-l-3}) well-defined.  The only price to be paid
for this transformation is an altered normalization, that will
cancel in the normalized density matrix (\ref{B-l-rho}).

    For a particular realization of the detector output $I(t')$, $0\leq t'\leq
t$, Eqs.\ (\ref{B-l-1})-(\ref{B-l-3}) define a linear map $\sigma
(0)\rightarrow \sigma (t)$, corresponding to a particular Kraus
operator $M_m$ (which, therefore, can be denoted as $M_{\{ I\}}$).
For the uncollapsing we have to realize the map, corresponding to
the inverse Kraus operator $C M_{\{ I\}}^{-1}$ (see Sec.\
\ref{general-theory}). It is obvious that in contrast to the case
of the non-evolving qubit, this cannot be done by simply continuing
the measurement and waiting for a specific result. The reason is
that now the map is characterized by 6 real parameters (8 parameters
for a linear operator $C M_{\{ I\}}^{-1}$ with neglected overall
phase and normalization), instead of 1 parameter for the
non-evolving case [see Eq.\ (\ref{murity}) and (\ref{Bayes-DD})]. We
will discuss a little later how the 6-parameter uncollapsing
procedure can be realized explicitly. Before that we discuss how to
find the operator $ M_{\{ I\}}$ in a more straightforward way, than
from Eqs.\ (\ref{B-l-1})-(\ref{B-l-3}).

    Let us consider only the evolution of (unnormalized) pure states
$|\psi (t)\rangle =\alpha (t) |1\rangle + \beta (t) |2\rangle$, so
that $\sigma = |\psi\rangle \langle \psi |$. Then Eqs.\
(\ref{B-l-1})-(\ref{B-l-3}) can be rewritten as
    \begin{eqnarray}
&& \dot{\alpha} = +i \frac{\varepsilon}{2}\alpha -i H \beta -\alpha
\frac{1}{2S_I} [I(t)-I_1]^2
 ,
\label{B-alpha}\\
&& \dot{\beta} = -i\frac{\varepsilon}{2}\beta - i H \alpha - \beta
\frac{1}{2S_I} [I(t)-I_2]^2,
     \label{B-beta}  \end{eqnarray}
where the infinite part of $I^2(t)$ can be canceled in the same way
as discussed above. The linearity of these equations guarantees that
for any given realization of $I(t)$, it is sufficient to solve
(\ref{B-alpha},\ref{B-beta}) for the initial states $|1\rangle$ and
$|2\rangle$ in order to find the solution for an arbitrary initial
state of the qubit.  Defining $\vec{v}_1 = \alpha_1(t) |1\ra +
\beta_1(t) |2\ra$ as the solution of (\ref{B-alpha},\ref{B-beta})
for initial state $|1\rangle$, and $\vec{v}_2 = \alpha_2(t) |1\ra +
\beta_2(t) |2\ra$ as the solution of (\ref{B-alpha},\ref{B-beta})
for initial state $|2\rangle$, we can write the solution for an
arbitrary initial state
$|\psi_{in}\rangle=|\psi(0)\rangle=a|1\rangle +b |2\rangle$ as
$|\psi(t)\rangle = a \vec{v}_1 +b \vec{v}_2$.
    Therefore, the Kraus operator $ M_{\{ I\}}$ for a given realization
of $I(t)$, in the $|1\ra, |2\ra$ basis is
     \be
M_{\{ I\} } = \left( \begin{array}{cc} \alpha_1 (t) \,\, &\alpha_2 (t) \\
\beta_1 (t) \,\, & \beta_2 (t)
\end{array} \right) .
    \label{M-I(t)}\ee

    For the uncollapsing we need to apply the Kraus operator $C
M_{\{ I\} }^{-1}$, which maps the state $\vec{v}_1$ onto
$C|1\rangle$ and the state $\vec{v}_2$ onto $C|2\rangle$. The reason
why we need a non-unitary transformation is that the vectors
$\vec{v}_1$ and $\vec{v}_2$ are in general non-orthogonal and have
different norms.  Geometrically, such a transformation can be done
by using a relative shrinking or stretching of two orthogonal axes
(found for a given $M_{\{ I\} }$), which would make $\vec{v}_1$ and
$\vec{v}_2$ orthogonal and equal in norm, followed by a unitary
transformation (this would correspond to the decomposition of the
form $U\sqrt{E}$ -- see Sec.\ \ref{formalism-wf}).  However, for a
practical realization of uncollapsing it is most natural to use the
shrinking or stretching of the axes $|1\rangle$ and $|2\rangle$ via
a continuous QND measurement with the QPC in the way considered
above for a non-evolving qubit. In this case the uncollapsing
procedure can be done in three steps (see Sec.\ \ref{formalism-wf}):
unitary evolution $V$, continuous QND measurement (where the qubit
Hamiltonian is turned off, $\varepsilon =H=0$) described by a
diagonal matrix $L$, and a final unitary operation $U$. (In the
notation of Sec.\ \ref{formalism-wf}, $V$ corresponds to
$V_L^\dagger U_m^\dagger$, and $U$ corresponds to $U_L^\dagger$.)
These operators should satisfy
    \be
    U L V  = C M_{\{ I\} }^{-1},
    \label{decompose} \ee
and it is easy to find $U$, $L$, and $V$ explicitly by recognizing
Eq.\ (\ref{decompose}) as a singular value decomposition of the
operator $C M_{\{ I\} }^{-1}$ (recall here that $L$ is diagonal;
also notice that the standard form for the singular value
decomposition is slightly different, with $V$ denoted as
$V^\dagger$).

    To find $L$ explicitly, we notice that
    \be
    C^{-2} M_{\{ I\}
}^\dagger  M_{\{ I\} } =UL^{-2}U^\dagger,
    \label{diagon-M_I}\ee
 which is simply the
diagonalization of $C^{-2} M_{\{ I\} }^\dagger  M_{\{ I\} }$.
Therefore,
    \be
    L = C \begin{pmatrix}
\lambda_-^{-1/2} & 0 \\ 0 & \lambda_+^{-1/2} \end{pmatrix} \quad
\mbox{or}
    \quad L = C \begin{pmatrix} \lambda_+^{-1/2} & 0 \\ 0 &
\lambda_-^{-1/2} \end{pmatrix}  ,
    \label{L-explicit} \ee
where
    \be \lambda_\pm = \frac{||\vec{v}_1||^2 + ||\vec{v}_2||^2}{2} \pm
\sqrt{\left(\frac{||\vec{v}_1||^2 - ||\vec{v}_2||^2}{2}\right)^2 +
|\vec{v}_1 \cdot \vec{v}_2^\ast|^2}
    \label{evals} \ee
 are the eigenvalues of the operator $M_{\{ I\}
}^\dagger  M_{\{ I\} }$ and the vectors $\vec{v}_i$ are defined
above Eq.\ (\ref{M-I(t)}). The Cauchy-Schwartz inequality,
$|\vec{v}_1 \cdot \vec{v}_2^\ast|^2 \le ||\vec{v}_1||^2
||\vec{v}_2||^2$, guarantees the non-negativity of $\lambda_-$. (The
notation $\vec{v}_1 \cdot \vec{v}_2^\ast$ is used for the inner
product $\langle v_2|v_1\rangle$ of $\vec{v}_2$ and $\vec{v}_1$).

    To find $U$, we use Eq.\ (\ref{diagon-M_I}) again and see that
the columns of $U$ are composed of the eigenvectors of $M_{\{ I\}
}^\dagger  M_{\{ I\} }$ [the sequence of columns depends on the
choice in Eq.\ (\ref{L-explicit})]. Finally, $V$ is given by
$V=U^\dagger CL^{-1}M^{-1}_{\{ I\} }$. For brevity we will not show
the matrices $U$ and $V$ explicitly.

    In the physical realization of the uncollapsing procedure the
measurement step $L$ can be performed in exactly the same way as in
Sec.\ \ref{uncollapsing-charge}. Comparing Eq.\ (\ref{L-explicit})
with Eqs.\ (\ref{Bayes-DD}) and (\ref{murity}), we see that the
continuous measurement by the QPC should be stopped when the
dimensionless measurement result $r(t)$ reaches the value \be
r_1=\ln \sqrt{\lambda_+/\lambda_-}>0 \quad \text{or}\quad r_2=\ln
\sqrt{\lambda_-/\lambda_+}<0, \label{rchoice} \ee for the first and
second choice in Eq.\ (\ref{L-explicit}), respectively (the choice
should be made beforehand, since it determines operation $V$). As
previously mentioned, the constant $C$ is not important here because
the physical state is always normalized. The procedure fails if the
desired result is not reached during the continuous measurement.

    The unitary operations $V$ and $U$ can be practically realized
in three substeps each: $z$-rotation on Bloch sphere by applying
non-zero energy asymmetry $\varepsilon$ for some time, $y$-rotation
by applying non-zero tunneling $H$, and then one more $z$-rotation.
However, the last $z$-rotation of $V$  and the first $z$-rotation of
$U$ are simply added to each other (since $L$ does not change the
relative phase of the state components or, equivalently, the azimuth
angle on the Bloch sphere). The corresponding trivial degree of
freedom can be eliminated, for example, by realizing the operation
$V$ in only two substeps, without the second $z$-rotation.

    Let us count the number of real parameters, characterizing the
uncollapsing procedure. Since $V$ and $U$ together provide $2\times
3-1=5$ parameters, and the desired result $r$ in the measurement
step adds one more parameter, the overall number of parameters is 6.
As expected, this is exactly the needed number of parameters
characterizing an arbitrary Kraus operator for the qubit (neglecting
normalization and overall phase). Let us also mention the fact from
linear algebra that the singular value decomposition
(\ref{decompose}) is unique in the non-degenerate case
($\lambda_+ > \lambda_- > 0$), up to the permutation of singular values
[corresponding to the choice in Eq.\ (\ref{L-explicit})] and
arbitrary phase factors in columns of $U$, with compensating changes
in $V$ (this corresponds to the discussed above compensation of
$z$-rotations).

    Now let us discuss the probability $P_S$ of the successful
uncollapsing. From the general theory discussed in section II [Eq.\
(\ref{P_S-wf})], it is bounded from above by a fraction, $P_S\leq
\min P_{\{I\}}/ P_{\{I\}} (|\psi_{in}\rangle )$, in which the
denominator is the probability density of the given realization
$I(t)$ for the initial state $|\psi_{in}\rangle = a|1\rangle
+b|2\rangle$ (with $|a|^2+|b|^2=1$), while the numerator is this
probability minimized over all initial states. So, the denominator
is given by the squared norm of the final state $|\psi(t)\rangle = a
\vec{v}_1+ b \vec{v}_2$,
    \be
    P_{\{ I\}} (|\psi_{in}\rangle ) = || a \vec{v}_1+ b \vec{v}_2 ||^2,
    \label{pofI} \ee
while the numerator is given by minimizing (\ref{pofI}) over all
normalized initial states. It is easy to see that this minimum is
equal to the minimum eigenvalue $\lambda_-$ of the operator $M_{\{
I\} }^\dagger  M_{\{ I\} }$, given by Eq.\ (\ref{evals}); therefore,
    \be
     P_S \le \frac{ \lambda_-}{|| a \vec{v}_1+ b \vec{v}_2 ||^2}.
    \label{P-S-ev-wf}\ee

    Converting this result into the language of density matrices and
simultaneously generalizing it to an arbitrary initial state
$\rho_{in}$, we obtain the bound
   \be
     P_S \le \frac{\frac{||\vec{v}_1||^2 + ||\vec{v}_2||^2}{2} -
\sqrt{\left(\frac{||\vec{v}_1||^2 - ||\vec{v}_2||^2}{2}\right)^2 +
|\vec{v}_1 \cdot \vec{v}_2^\ast|^2}}{\rho_{in,11} ||\vec{v}_1||^2 +
\rho_{in,22} ||\vec{v}_2||^2 + 2
\mbox{Re}[\rho_{in,12}\,\vec{v}_1\cdot \vec{v}_2^\ast ] } ,
    \label{P-S-ev-dm}\ee
in which the numerator is the explicit expression (\ref{evals}) for
$\lambda_-$. It is easy to check that this result reduces to the
bound (\ref{undo-general}) in the non-evolving case, in which
$\vec{v}_1 = \left( \sqrt{P_1 (\bar{I})}, 0\right)^T$ and $\vec{v}_2
= \left( 0, \sqrt{P_2(\bar{I})}\right) ^T$.

    The uncollapsing procedure discussed in this subsection
is optimal in the sense that it corresponds to the upper bound of
Eq.\ (\ref{P-S-ev-dm}). To prove this statement, instead of
calculating $P_S$ explicitly, let us use the fact (see Sec.\
\ref{sect-informat}) that the product $P_S   P_{\{ I\}}$ cannot
depend on the initial state.  Therefore, it is sufficient to prove
the optimality of $P_S$ only for one initial state. Let us choose
the state $|\psi_{in}\rangle$ that is the eigenvector of $M_{\{ I\}
}^\dagger  M_{\{ I\} }$, corresponding to the eigenvalue
$\lambda_-$. Then after the first measurement (operator $M_{\{ I\}
}$) and the unitary operation $V$ it is transformed into one of the
basis states [$|1\rangle$ or $|2\rangle$ for the first or second
choice in (\ref{L-explicit}), respectively].  Recall that for the
QND (non-evolving) measurement case, $r(t) \rightarrow \infty$ for
the initial state $|1\ra$, while $r(t) \rightarrow -\infty$ for the
initial state $|2\ra$. The crossing thresholds (\ref{rchoice})
indicate that the measurement $L$ is always successful because
$r(t)$ necessarily crosses the desired value (which is positive for
$|1\rangle$ and negative for $|2\rangle$, as discussed above).
Therefore, the uncollapsing success probability for this special
state is 100\%, that is equal to the upper bound (\ref{P-S-ev-dm}).
As mentioned above, the optimality of the procedure for this special
state also proves the optimality for any initial state.

    Obviously, an uncollapsing procedure can also be non-optimal. As
an example, let us consider a procedure which realizes the desired
mapping $\{ \vec{v}_1, \vec{v}_2 \} \rightarrow \{ C|1\rangle, C
|2\rangle \}$ using two measurements instead of one. The goal of the
first measurement is to map $\{ \vec{v}_1, \vec{v}_2 \}$ into an
orthogonal pair of vectors, while the goal of the second measurement
is to equalize their norms, keeping them orthogonal.
    The first goal can be achieved by stretching/shrinking of the
Hilbert space along any axis $\vec{u}$ of the form $\vec{v}_1+c
\vec{v}_2 (\vec{v}_1\cdot \vec{v}_2)/|\vec{v}_1\cdot \vec{v}_2|$
with an arbitrary positive real number $c$ (it is easy to visualize
this procedure of making two vectors orthogonal by assuming the
space of real vectors, for which the axis $\vec{u}$ is geometrically
in between $\vec{v}_1$ and $\vec{v}_2$; the same geometrical idea
works for complex vectors). We recall that measurement for a
non-evolving qubit (Sec.\ \ref{charge-dynamics}) stretches
(squeezes) the $|1\ra$ axis, while squeezing (stretching) the
$|2\ra$ axis. Therefore, the first goal can be achieved by a unitary
operation which rotates $\vec{u}$ into $|1\ra$, followed by a
continuous measurement (with a QPC) of a non-evolving qubit, to be
stopped when the mapped vectors become orthogonal. After the vectors
$\{ \vec{v}_1, \vec{v}_2 \}$ are transformed into an orthogonal pair
by the (successful) first measurement, the second part of the
procedure should stretch/shrink the 2D Hilbert space along the
resulting vectors to make them equal in norm. This can be done
similarly, by a unitary rotation and partial measurement of a
non-evolving qubit. Finally, another unitary operation can be used
to map the resulting pair of vectors into $\{ C|1\rangle, C
|2\rangle \}$, thus completing the uncollapsing procedure. Notice
that both measurements are performed in the ``wait and stop''
manner, and both measurements should be successful to realize the
uncollapsing. While the successfully uncollapsed state is still
perfect in this procedure, the probability of success is lower than
the bound (\ref{P-S-ev-dm}). To prove this non-optimality, let us
again use the initial eigenstate $|\psi_{in}\rangle$, which corresponds
to eigenvalue $\lambda_-$, so that the bound (\ref{P-S-ev-dm}) is 100\%. Then
the probability of success for the first measurement is in general
less than 100\% (it is 100\% only for one specific axis discussed
previously, while here we consider a range of possible axes by
allowing $c$ to vary). Thus, the success probability is less than
100\% for this special state, and therefore $P_S$ is below the bound
(\ref{P-S-ev-dm}) for any initial state.

\section{Phase qubit}

\begin{figure}
\centerline{ \epsfxsize=2.7in \hspace{0.3cm} \epsfbox{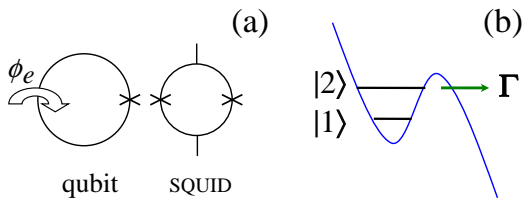} }
\vspace{0.3cm} \caption{(Color online).  (a) Schematic of
a phase qubit controlled by an external flux $\phi_e$ and
inductively coupled to the detector SQUID. (b) Energy profile
$V(\phi )$ with quantized levels representing qubit states. The
tunneling event through the barrier is sensed by the SQUID.}
\label{fig-phase}\end{figure}

The second explicit example of erasing information and uncollapsing
the wavefunction is for a superconducting phase
qubit.\cite{Martinis} The system is comprised of a superconducting
loop interrupted by one Josephson junction [Fig.\
\ref{fig-phase}(a)], which is controlled by an external flux
$\phi_e$ in the loop. Two qubit states $|1\rangle$ and $|2\rangle$
[Fig.\ \ref{fig-phase}(b)] correspond to two lowest states in the
quantum well for the potential energy $V(\phi)$ where $\phi$ is the
superconducting phase difference across the junction. (Notice that
the standard notation for the phase qubit states is $|0\rangle$ and
$|1\rangle$;\cite{Martinis} however, we use $|1\rangle$ and
$|2\rangle$ for consistency with the previous section.)
 The qubit is
measured by lowering the barrier (which depends on $\phi_e$), so
that the upper state $|2\rangle$ tunnels into the continuum with the
rate $\Gamma$, while the state $|1\rangle$ does not tunnel out. The
tunneling event is sensed by a two-junction detector SQUID
inductively coupled to the qubit [Fig.\ \ref{fig-phase}(a)].
Transitions between the levels $|1\rangle$ and $|2\rangle$ can be
induced by applying microwave pulses that are resonant with the
energy level difference.

\subsection{Partial collapse}

    For sufficiently long tunneling time $t$, $\Gamma t \gg 1$, the measurement
is a (partially destructive) projective measurement: \cite{Martinis}
the system is destroyed if the tunneling occurs,  while if there is
no record of tunneling, then the state is projected onto the lower
state $|1\ra$.  This measurement technique is remarkable in the fact
that the wavefunction is collapsed {\it if nothing happens}. A more
subtle situation arises if the barrier is raised after a finite time
$t \sim \Gamma^{-1}$: then the measurement is only partial and
therefore is of a POVM-type.\cite{Katz} The system is still
destroyed if tunneling happens, while in the case of no tunneling
(which we refer to as a null-result measurement) the state is
partially collapsed. This situation may be described with a
two-outcome POVM, with elements $E_n$ and $E_y$, where $n$ denotes
the null result, and $y$ denotes the affirmative (tunneling) result.
The POVM elements, given in the $|1\ra$, $|2\ra$ basis are
    \be
E_n =\begin{pmatrix} 1 & 0 \\ 0 & e^{-\Gamma t}
\end{pmatrix}, \qquad E_y =\begin{pmatrix} 0 & 0 \\ 0  &
1 - e^{-\Gamma t}\end{pmatrix}, \label{POVM-phase} \ee
    with the obvious completeness relation $E_n+E_y={\bf 1}$.

    It is interesting to notice that while $E_n=M_n^\dagger M_n$
corresponds to a Kraus operator $M_n$ (see discussion below), no
meaningful Kraus operator $M_y$ can be introduced for the POVM
element $E_y$, because in the case of a tunneling event the system
leaves its two-dimensional Hilbert space and becomes incoherent (so
that a single Kraus operator cannot be introduced even in the
extended Hilbert space). However, this is not important for us
because we are interested in the null-result case only.

    Limiting the unitary operation in the decomposition (\ref{mop})
to be the phase factor only, we may expect the null-result Kraus
operator $M_n$ to be of the form $M_n={\rm diag}\{1,e^{-\Gamma t/2}
e^{-i\varphi}\}$. In a simple model \cite{Dalibard,Pryadko} the
phase $\varphi$ is zero in the rotating frame, which compensates for
the energy difference of the states $|1\rangle$ and $|2\rangle$. In
a real experiment,\cite{Katz} however, this energy difference
changes in the process of measurement because it is affected by
$\phi_e$, and therefore even in the rotating frame the phase
$\varphi$ is non-zero. Correspondingly, the qubit density matrix
changes after the null-result measurement as
    \begin{equation}
\frac{\rho_{11}(t)}{\rho_{22}(t)}= \frac{\rho_{in,11}}{ e^{-\Gamma
t}\rho_{in,22}} , \,\,\,
\frac{\rho_{12}(t)}{\sqrt{\rho_{11}(t)\rho_{22}(t)}} =
\frac{e^{i\varphi
(t)}\rho_{in,12}}{\sqrt{\rho_{in,11}\rho_{in,22}}}.
    \label{Bayes-phase}
    \end{equation}
In the real experiment\cite{Katz} the situation is even more complex
because the tunneling rate gradually changes in time; also, instead
of controlling the measurement time $t$, it is much easier to
control the tunneling rate. As a result, the measurement should be
characterized by the overall strength
$p_t=1-\exp[-\int_0^t\Gamma(t')dt']$. Nevertheless, for simplicity,
we use here the physically transparent language of Eq.\
(\ref{Bayes-phase}) with $e^{-\Gamma t}$ understood as $1-p_t$.

Up to such changes of notation, the coherent non-unitary evolution
(\ref{Bayes-phase}) has been experimentally verified in Ref.\
\onlinecite{Katz} using tomography of the post-measurement state.
The state tomography consisted of 3 types of rotations of the qubit
Bloch sphere, followed by complete (projective) measurement.
 Actually, in the experiment it was not possible to
select only the null-result cases, because it was not possible to
distinguish if a tunneling event happened during measurement or
during tomography. However, a simple trick of comparing the
protocols with and without tomography made it possible to separate the
null-result cases.

Notice that except for the effect of extra phase $\varphi (t)$, the
qubit evolution (\ref{Bayes-phase}) is similar to the qubit
evolution in the example of Sec.\ \ref{charge-dynamics}; in
particular, it also represents an ideal measurement which does not
decohere the qubit. Formally, the evolution (\ref{Bayes-phase})
corresponds to the measurement result $r=\Gamma t/2$ in Eq.\
(\ref{Bayes-DD}). As will be shown later, the probability to undo
the measurement is still given by Eq.\ (\ref{s}) using this value of
$r$.

\subsection{Uncollapsing}

We will now describe how to undo the state disturbance
(\ref{Bayes-phase}) caused by the partial collapse resulting from
the null-result measurement. The undoing of this measurement
consists of three steps:  $(i)$ Exchange the amplitudes for the
states $|1\rangle$ and $|2\rangle$ by application of a microwave
$\pi$-pulse, $(ii)$ perform another  measurement by lowering the
barrier, identical to the first measurement, $(iii)$ apply a second
$\pi$-pulse. If the tunneling did not happen during the second
measurement, then the information about the initial qubit state is
canceled (both basis states have equal likelihood for two
null-result measurements). Correspondingly, according to Eq.\
(\ref{Bayes-phase}) (which is applied for the second time with
exchanged indices $1\leftrightarrow 2$), any initial qubit state is
fully restored.  An added benefit to this strategy is that the phase
$\varphi$ is also canceled automatically; the physics of this
{\it phase cancellation} is the same as in the spin-echo technique for qubits.

It is easy to mistake the above pulse-sequence as simply the well
known spin-echo technique alone.  We stress that this is not the
case:  Spin-echo deterministically reverses an unknown unitary
transformation (arising usually from a slowly varying magnetic
field) without gaining or losing any information about what that
state is.  Our strategy is probabilistic and
requires erasing the classical information that one extracts from
the system to begin with.  It is a (probabilistic) reversal of a
known non-unitary transformation - and therefore quite different
from spin echo.

    The success probability $P_S$ for the uncollapsing strategy
described above may be calculated by noting that it is just the
probability that the second measurement gives a null result.  If we
start with the qubit state $\rho_{in}$, the state after the first
measurement is given by Eq.\ (\ref{Bayes-phase}), and after the
$\pi$-pulse the occupation of the upper state is $\tilde{\rho}_{22}
= \rho_{in,11}/[\rho_{in,11}+\rho_{in,22}\,e^{-\Gamma t}]$.  The
success probability is simply the probability that the second
tunneling will not occur, $P_S= \tilde{\rho}_{11} +
\tilde{\rho}_{22} e^{-\Gamma t} =  1-\tilde{\rho}_{22} (1-e^{-\Gamma
t})$, which can be expressed as
    \begin{equation}
P_S=\frac{e^{-\Gamma t}}{\rho_{in,11}+e^{-\Gamma t}\rho_{in,22}},
    \label{s-phase}
    \end{equation}
and formally coincides with Eq.\ (\ref{s}) for $r=\Gamma t/2$. We
can verify that this strategy is optimal by using the $E_n$ POVM
element (\ref{POVM-phase}) together with the general result
(\ref{gen-result}) to find the upper bound for the success
probability $P_S$.  The numerator of (\ref{gen-result}) is the
smallest eigenvalue of $E_n$, which is $\exp(-\Gamma t)$, while the
denominator is ${\rm Tr} E_n \rho_{in} = \rho_{in,11}+e^{-\Gamma
t}\rho_{in,22}$, giving a $P_S$ that coincides with Eq.\
(\ref{s-phase}), and thus confirming the optimality of the analyzed
uncollapsing procedure.

    The total uncollapsing probability $\tilde{P}_S$ of two null
results (see Sec.\ II\,D) is
    \begin{equation}
    \tilde{P}_S=(\rho_{in,11}+e^{-\Gamma t}\rho_{in,22}) P_S = e^{-\Gamma
    t}.
    \end{equation}
As expected (see Sec.\ II\,D) this probability does not depend on
the initial state.


Uncollapsing of the phase qubit state has recently been
experimentally realized by Nadav Katz and colleagues in the lab of
John Martinis, at UC Santa Barbara.\cite{Katz2} The experimental
protocol was slightly shorter than that described above: it was
missing the second $\pi$-pulse, so the uncollapsed state was
actually the $\pi$-rotation of the initial state. Shortening of the
protocol helped in decreasing the duration of the pulse sequence,
which was about 45 ns, including the state tomography. Since the
qubit energy relaxation and dephasing times were significantly
longer, $T_1=450$ ns and $T_2^*=350$ ns, the simple theory described
above was sufficiently accurate. The same trick as for the
partial-collapse experiment \cite{Katz} was used to separate
tunneling events during the first, second, and tomography
measurements, because the detector SQUID was too slow to distinguish
them directly.

The uncollapse procedure should restore any initial state. However,
instead of examining all initial states to check this fact, it is
sufficient to choose 4 initial states with linearly independent
density matrices and use the linearity of quantum
operations.\cite{Nielsen} In the experiment \cite{Katz2} the
uncollapse procedure was applied to the initial states
$(|1\ra+|2\ra)/\sqrt{2}$, $(|1\ra- i |2\ra)/\sqrt{2}$,  $|1\ra$, and
$|2\ra$, and then the results were expressed via the language of the
quantum process tomography \cite{Nielsen,QPT-note} (QPT). The
experimental \cite{Katz2} QPT fidelity of the uncollapsing procedure
was above 70\% for $p_t < 0.6$. A significant decrease of the
uncollapsing fidelity for larger measurement strength $p_t$,
especially for $p_t>0.8$, was due to finite $T_1$ time and the fact
that the null-result selection preferentially selects the cases with
energy relaxation events, so that the procedure should no longer
work well when $1-p_t$ becomes comparable to the probability of
energy relaxation.

    As mentioned above, the uncollapsing procedure described in this
subsection is theoretically optimal in the sense that it maximizes
the bound (\ref{gen-result}) for the uncollapsing probability. An
example of a non-optimal uncollapsing for a phase qubit was
considered in Ref.\ \onlinecite{Ruskov-07}. It was shown that if the
measurement process is performed simultaneously with Rabi
oscillations, then in the null-result case the initial state is
periodically restored. The non-optimality of uncollapsing for such a
procedure is due to measurement of an evolving qubit, which
corresponds to a sequence of many measurements; a similar reason for
the non-optimality of the two-step uncollapsing was discussed at the
end of Sec.\ \ref{charge-evolve}.

    Exact uncollapsing requires an ideal detection, which does not
decohere a quantum state; Eq.\ (\ref{Bayes-phase}) corresponds to
such an ideal detection. However, if various decoherence mechanisms
are taken into account, \cite{Pryadko} then only imperfect
uncollapsing is possible. The theory of imperfect uncollapsing is a
subject of further research.

\section{General procedure for entangled charge qubits}

Let us present an explicit procedure which can be used in principle
to undo an arbitrary measurement $M_m$ of any number $N$ of
entangled qubits with maximum probability.  For simplicity we
consider double-quantum-dot charge qubits and assume that any
unitary transformation can be used in the procedure. If the operator
$M_m$ was produced by a one-qubit measurement, and other entangled
qubits were not experiencing a Hamiltonian evolution, then the
formalism of Sec.\ \ref{charge-dynamics} is essentially unchanged,
\cite{Kor-ent} and uncollapsing of the measured qubit leads to the
restoration of the whole entangled state. If the operator $M_m$ was
produced by a one-qubit measurement, while other qubits were
evolving in a unitary way but not interacting with the measured
qubit, then uncollapsing is also easy: we should uncollapse the
measured qubit in the usual way (Sec.\ III) and should apply inverse
unitary transformation for other qubits. In this section, however,
we do not consider these simple special cases; the goal is to undo
an arbitrary Kraus operator $M_m$.

    Let us decompose $M_m$ as $M_m=U_m\sqrt{E_m}$ [see Eq.\
(\ref{mop})]. Reversing the unitary operation $U_m$ can be done in
the regular Hamiltonian way, so the nontrivial part is undoing the
$\sqrt{E_m}$ operator.  We recall the diagonalization of $E_m$ is
given by $E_m = \sum_i p^{(m)}_i |i\ra \la i |$ with vectors
$|i\rangle$ forming an orthonormal basis. As discussed in
Sec.~II\,B, for the optimal uncollapsing which maximizes the success
probability, we have to perform a procedure corresponding to another
measurement operator $\tilde{L}=\sqrt{\min_j p_j^{(m)}} \,
E_m^{-1/2}$ which is also diagonal in the basis $|i\rangle$ with
corresponding matrix elements $\tilde{L}_{ii} = \la i |\tilde{L}
|i\rangle=\sqrt{(\min_j p_j^{(m)})/p_i^{(m)}}$, all of which are
between 0 and 1. Given $N$ qubits, $i$ ranges from 1 to $2^N$.
Notice that $\tilde{L}$ is obviously Hermitian.

\begin{figure}
\centerline{ \epsfxsize=6cm \epsfbox{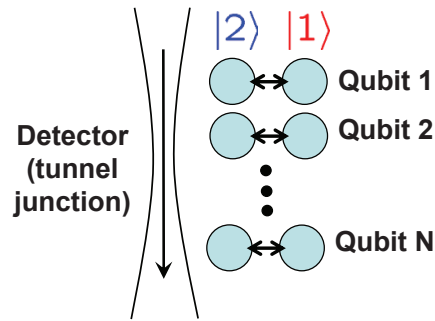} }
\caption{(Color online.)  Schematic set-up for uncollapsing of $N$
entangled qubits.  The tunnel junction detector (QPC) is in a
strongly nonlinear regime, so that an electron can tunnel through it
with rate $\gamma$ only when all qubits are in state $|1\ra$. }
\label{fig-entangled}
\end{figure}

Our procedure is to realize $\tilde{L}$ with a sequence of
null-result measurements and unitary operations. Shown in
Fig.~\ref{fig-entangled} is an illustration of the physical set-up
that is used for the measurements: a QPC (tunnel junction)
capacitively coupled to $N$ non-evolving DQD charge qubits. We
assume that the QPC is tuned to a highly nonlinear regime, for which
no electron can tunnel across the QPC barrier on experimentally
relevant time-scales unless all qubits are in the state $|1\ra$.  We
name this multi-qubit state $|\openone \ra \equiv |1,1,\ldots,
1\ra$. Such a regime is possible because of the exponential
dependence of the tunneling rate on QPC barrier height, while the
barrier height depends linearly on the states of the coupled qubits.
Of course, this regime is not quite realistic; however, we discuss
the procedure in principle.
  We also assume that even for the $N$-qubit state $|\openone \ra$,
the rate $\gamma$ of electron tunneling through the QPC is rather
low, so that we can distinguish single tunneling events
(technically, this would require an additional single-electron
transistor). If we perform the measurement during time $t$ and see
no tunneling through the QPC, then similarly to the case of Sec.\
IV\,A, the corresponding null-result Kraus operator $M_n$ shrinks
the $|\openone \ra$ axis of the Hilbert space by the factor
$e^{-\gamma t/2}$, while leaving all perpendicular axes unchanged.
For the matrix elements this means $\la \openone | M_n |\openone \ra
= e^{-\gamma t/2}$, $\la\psi_j^\perp| M_n |\psi_{j'}^\perp \ra =
\delta_{jj'}$, $\la\psi_j^\perp| M_n |\openone \ra = \la \openone |
M_n | \psi_j^\perp \ra = 0$, where we introduced a set of $2^N -1$
states $|\psi_j^\perp\ra$ spanning the subspace orthogonal to $|
\openone\ra$.

The general strategy to implement the operator $\tilde{L}$ is the
following.   We first note that in the basis $| i \ra$ that
diagonalizes $\tilde{L}$, this diagonal matrix can be represented as
a product of $2^N$ diagonal matrices, where each term in the product
has all diagonal entries as 1, except the $i$th entry: ${\rm diag}\{
1, 1, \ldots, \tilde{L}_{ii}, \ldots, 1\}$.  Each of these matrices
may be interpreted as a separate Kraus operator that can be
sequentially implemented. Thus, the explicit physical procedure
consists of $2^N$ steps, each of which has 3 substeps.  First, we
apply a unitary transformation $U_1$ which transforms the first
basis vector $|i=1\rangle$ into the state $|\openone \rangle$. Then
the evolution of all qubits is stopped, and the detector is turned
on for a time $t_1$. This time is chosen so that the null-result
Kraus operator $L^{(1)}$ has the desired matrix element $\la
\openone| L^{(1)}|\openone \ra= \tilde{L}_{11}$; this condition
yields $t_1 = -2\gamma^{-1} \ln \tilde{L}_{11}$.  The
measurement is then followed by the reverse unitary, $U_1^\dagger$,
to take the state $|\openone \ra$ back to state $|i=1\ra$.  This
3-substep procedure is then repeated for $i=2, 3, \ldots, 2^N$,
sequentially transforming the state $|i\ra$ to $|\openone \ra$ with
unitary $U_i$, and performing measurement with the detector for a
time $t_i = -2\gamma^{-1} \ln \tilde{L}_{ii}$, followed by the
reverse unitary, $U_i^\dagger$.  This sequence of
steps decomposes the uncollapsing operator $\tilde{L}$ as
    \be \tilde{L} = U_{2^N}^\dagger L^{(2^N)}
U_{2^N} \cdots U_2^\dagger L^{(2)} U_2 U_1^\dagger L^{(1)} U_1. \ee

The uncollapsing procedure is successful only if there were no
tunneling events in the QPC. By construction, the success
probability $P_S$ for this procedure maximizes the general bound
(\ref{gen-result}).

    The success probability $P_S$ for the uncollapsing process
$\rho_m \rightarrow \rho_{in}$ with
$\rho_{in}=\tilde{L}\tilde{\rho}\tilde{L}^\dagger
/\mbox{Tr}(\tilde{L}^\dagger \tilde{L} {\tilde \rho})$ and
$\tilde{\rho}=U_m^\dagger \rho_m U_m$, can be calculated as
 \be P_S = \mbox{Tr}(\tilde{L}^\dagger \tilde{L} {\tilde \rho})= \sum_i
\tilde{L}_{ii}^2 \tilde{\rho}_{ii} =\sum_i \tilde{\rho}_{ii}
\exp(-\gamma t_i),
    \label{success} \ee
where ${\tilde \rho}_{ij}$ are the matrix elements of $\tilde{\rho}$
in the basis $|i\ra$, which diagonalizes $\tilde{L}$. We may also
find this result from another perspective by realizing that the
success probability is simply the product of the null-result
probabilities $p_S^{(i)}$ of all $2^N$ measurements,
    \be
    P_S=\prod_i^{2^N} p_S^{(i)}, \,\,\, p_S^{(i)}=
\frac{\sum_{j=1}^{i}\tilde{\rho}_{jj}e^{-\gamma
t_j}+\sum_{j=i+1}^{2^N}\tilde{\rho}_{jj}
}{\sum_{j=1}^{i-1}\tilde{\rho}_{jj}e^{-\gamma
t_j}+\sum_{j=i}^{2^N}\tilde{\rho}_{jj}} ,
    \label{P_S&p_S}\ee
where the expression for $p_S^{(i)}$ comes from comparing the traces
of unnormalized density matrices after each of $2^N$ steps of the
procedure.
    It is instructive to show explicitly that this expression for
$p_S^{(i)}$ is equal to the expected expression
    \be
    p_S^{(i)}=1-(1-e^{-\gamma t_i}) \tilde{\rho}_{ii}^{(i)},
    \label{p_S-norm}\ee
in which $\tilde{\rho}^{(i)}$ is the normalized density matrix
before the $i$th step of the procedure (after $i-1$ null-result
steps).  This can be done if we prove the relation
  \be \prod_{i=1}^k
[1-\tilde\rho_{ii}^{(i)}(1-e^{-\gamma t_i})]=1-\sum_{i=1}^k
(1-e^{-\gamma t_i})\tilde\rho_{ii},
    \label{prod=1-sum}\ee
    [notice that the right-hand-side of this equation is equal to the numerator
    in Eq.\ (\ref{P_S&p_S}) with substitution $k\rightarrow i$].
Equation (\ref{prod=1-sum}) can be proven by induction using the
relation
$\tilde{\rho}_{ii}^{(i)}=\tilde{\rho}_{ii}/
 \prod_{j=1}^{i-1}[1-(1-e^{-\gamma
t_j})\tilde\rho_{jj}^{(j)}]$, which can be easily derived
recursively, $\tilde{\rho}_{ii}^{(j)}\rightarrow
\tilde{\rho}_{ii}^{(j+1)}$, starting from
$\tilde\rho_{ii}^{(1)}=\tilde\rho_{ii}$.  In this way we show
consistency between the null-result probabilities given by Eqs.\
(\ref{P_S&p_S}) and (\ref{p_S-norm}), permitting the calculation of $P_S$ in
two independent ways.

    Let us mention again that the uncollapsing procedure considered
in this section reaches the upper bound (\ref{gen-result}) for the
success probability $P_S$, that can be seen both by construction and
explicitly.

\section{Recent developments in wavefunction uncollapse}
Before concluding, we wish to give a summary of some interesting
recent  developments in this area of research.   We will briefly
discuss one theory proposal and one experiment.

\subsection{Spin qubit}
The examples given above mainly concern quantum dot charge qubits.
It is a natural question if a similar kind of partial
collapse/uncollapse can be carried over to spin qubits.  An analysis
of this situation was carried out by Trauzettel, Burkard, and one of
the authors.\cite{spinwm}  There it was shown how an
uncollapse measurement can be realized using a scheme similar to the
recent experiments by Koppens {\it et al.}\cite{leo}  The essential
idea of the spin-qubit experiments
\cite{leo,nuclear_exp,hanson,elzerman} is to manipulate and measure
the spin of a single electron through the charge degree of freedom.
This technique circumvents the otherwise difficult problem of
controlling the weakly interacting spin.  While we refer the reader
to Refs.~\onlinecite{spinwm,leo,nuclear_exp,hanson,elzerman} for the
details, we will give a simplified thumb-nail sketch of the physics
here.

 The qubit is encoded with two electron spins, where each electron is
confined in a separate quantum dot.  In contrast to our charge qubit
discussion, these dots are open, with electrons able to enter and
leave.  Electrical bias is applied across this double quantum dot
leading to charge transport.  Electrons can tunnel sequentially, but
{\it spin blockade}\cite{Ono} restricts transport to situations
where the two electrons form a spin singlet $(0,2)S$ on the right
dot while the spin triplet $(0,2)T$ is outside the transport energy
window due to the large single quantum dot exchange energy [here
$(n, m)$ refers to $n$ electrons on the left dot and $m$ electrons
on the right dot].   This blockade physics provides an interesting
initialization procedure of the quantum register - when the
single-electron current stops flowing, we are confident that the
two-electron state is in a $(1, 1)T$ state, because in the absence
of spin flip processes, the tunneling transition to the $(0, 2)$
state is forbidden.   From this configuration, it is possible to
manipulate the system by applying electron spin resonance
pulses,\cite{leo} transitioning the state to have overlap with the
singlet state.   Thus, the electron on the left dot may tunnel (with
rate $\Gamma$) to the right dot and exit the system, giving rise to
a small electrical current at the drain when this process is
repeated many times.  Of course, this will happen with some
probability controlled by the overlap of the state with the singlet.

 Drawing on our experience with the phase qubit, it is clear how to
devise a weak measurement experiment and an uncollapsing experiment:
the allowed transition can be permitted for a time of one's choosing
and then forbidden by detuning the energy levels with a voltage
pulse to one of the quantum dot's gates. In this way one can weakly
probe the two-electron state, and in the null-result case (no single
electron tunneling) partially collapse it to the  triplet subspace.
In order to propose the uncollapse part of the experiment, it is
easiest to consider the case when the nuclear spins quickly admixed
the singlet state with a triplet state, permitting the two-qubit
state to encode one effective qubit: parallel or anti-parallel
spins.  The weak measurement technique described above will then
partially collapse the state toward the parallel state under a
null-measurement (no single electron tunneling).  If now a
$\pi$-pulse is applied to one of the spins with electron spin
resonance, followed by a second null-measurement, this was shown to
uncollapse the state of the effective qubit.\cite{spinwm}

\subsection{Optical polarization qubit}
Another interesting development is the experimental implementation
of wavefunction uncollapse for optical qubits using the polarization
degree of freedom of single photons by Kim {\it et al.}\cite{kim2009}  The weak
measurement was implemented by passing the photon through a glass
plate oriented at the Brewster angle.  Only the vertical
polarization is reflected off of the glass plate (with some probability).
By placing a single-photon detector where the
photon would have gone had it reflected, a null-click measurement
partially collapses the polarization state to the horizontal
polarization. The strength of the measurement can be increased by
placing a series of plates in a row, effectively increasing the net
probability of a vertically-polarized photon reflecting at some
point.

The wavefunction uncollapse is done by inserting a half-wave plate
(exchanging the amplitude of horizontal with vertical polarization),
and having the same number of plates traversed by the photon again.
If none of the single-photon detectors click, the polarization state
is uncollapsed.  This has been verified \cite{kim2009} with quantum
state tomography (with polarizer and single-photon counter placed
after all of the reflecting plates) on the photon, conditioned on
none of the other photon detectors firing.  The experiment showed an
uncollapsing fidelity of above $94\%$ for measurement strengths up
to 0.9. It was also pointed out that the information from the first
weak measurement  can be used for developing guessing strategies
about the unknown initial state.  Two such strategies were
presented, and one was shown to be optimal.  Of course, in the case
where the measurement was subsequently undone, these strategies did
no better than random guessing.

\section{Conclusion}
We have reviewed and extended recent developments in the theory (and
experiment) of wavefunction uncollapse by undoing quantum
measurements.  We have formulated the problem of
wavefunction uncollapse in terms of a contest
between the uncollapse proponent, Plato, and the uncollapse skeptic,
Socrates, monitored by the arbiter, Aristotle. Plato claims to have
the ability to uncollapse wavefunctions, and this ability can be
tested under the rules of the contest set forth. 

 We have discussed several general
features of the uncollapse process in the abstract case, such as the
upper bound on the success probability and quantum information
aspects of the problem.  In order to probabilistically undo the
measurement, it is necessary to erase the information extracted
about the state in the first measurement. This is a necessary
condition to uncollapse the wavefunction, because otherwise various
paradoxes arise. However, the information erasure is surely not a
sufficient condition: the unitary evolution should also be properly
reversed and, as the most experimentally challenging condition, the
process should not bring decoherence, which requires a very good
(ideal) detector.

 In addition to discussing the theory of
wavefunction uncollapse in the abstract case, we have also considered
a variety of solid-state implementations and
specific practical strategies for wavefunction uncollapse.  Not only the success probability, these specific systems also allow
the calculation of other characteristics of the process, such as the
waiting time distribution for wavefunction uncollapse in the charge
qubit case. The cases of the charge qubit (with and without
Hamiltonian dynamics), the phase qubit, and many entangled charge
qubits have been examined in detail.    Additionally, we have also
discussed two experimental realizations of this physics, based on
the phase qubit and the polarization qubit, both of which have
clearly demonstrated wavefunction uncollapse with high fidelity.

The ideality of the detector is necessary for
perfect uncollapsing, and we have only dealt with these kinds of
detectors in the theory section of this paper (by this we mean the
detector adds no extra decoherence to the system). If a detector is
slightly non-ideal, then even a perfectly executed uncollapse
strategy will result in a slight infidelity in the final state.
This is indeed the case in the experiments mentioned above although
the fidelity was quite high.  In such a situation there are two
characteristics to contend with: the fidelity of uncollapsing as
well as the probability of claimed success. It is an open topic for
future research how these characteristics are related.


\begin{acknowledgments}
The work was supported by NSA and ARDA under ARO grant
W911NF-08-1-0336, the National Science Foundation under
grant No. DMR-0844899, and the University of Rochester.
\end{acknowledgments}


\begin{thebibliography}{99}
\bibitem{Neumann} J. von Neumann, {\it Mathematical Foundations of
        Quantum Mechanics\/} (Princeton Univ. Press, Princeton, 1955).

\bibitem{wheeler}
J. A. Wheeler,  {\it Law Without Law}, in {\it Quantum Theory and
Measurement},  Edited by J. A. Wheeler and W. H. Zurek, (Princeton
Series in Physics, Princeton University Press, 1983).

\bibitem{Davies-Kraus-Holevo} E. B. Davies, {\it Quantum Theory of Open
    Systems} (Academic, London, 1976); K. Kraus, {\it States, Effects,
    and Operations: Fundamental Notions of Quantum Theory} (Springer, Berlin,
    1983);  C. M. Caves, Phys. Rev. D {\bf 33}, 1643 (1986);
H. J. Carmichael, {\it An Open System Approach to Quantum Optics},
Lecture notes in physics (Springer, Berlin,  1993);
M. B. Mensky, Phys. Usp. {\bf 41}, 923 (1998); A.
S. Holevo, {\it Statistical Structure of Quantum theory} (Springer,
2001); D. F. Walls and G. J. Milburn, {\it Quantum Optics},
(Springer, 2006).

\bibitem{us-undo}
A. N. Korotkov and A. N. Jordan, Phys. Rev. Lett. {\bf 97},
166805 (2006).

\bibitem{us-proceeding}
A. N. Jordan and A. N. Korotkov, in {\it Coherence and Quantum
Optics IX}, edited by N. P. Bigelow et al. (Optical Soc. of America,
2008), p. 191.


\bibitem{Ueda-99}
M. Koashi and M. Ueda, Phys. Rev. Lett. {\bf 82}, 2598
    (1999).

\bibitem{others} A. Ban, J. Phys. A {\bf 34}, 9669 (2001); G.M. D'Ariano,
    Fortschr. Phys. {\bf 51}, 318 (2003); M. A. Nielsen and C. M. Caves, Phys. Rev. A {\bf 55}, 2547
    (1997); A. Royer, Phys. Rev. Lett. {\bf 73}, 913 (1994); {\bf 74},
    1040(E) (1995); M. Ueda, N. Imoto, and H. Nagaoka, Phys. Rev. A {\bf 53},
    3808 (1996); H. Mabuchi and P. Zoller, Phys. Rev. Lett. {\bf 76}, 3108
    (1996); F. Buscemi, M. Hayashi, M. Horodecki, Phys. Rev. Lett. {\bf 100}, 210504 (2008).

\bibitem{gefter}
For various perspectives on these
developments, see A. Gefter, New Scientist {\bf 194}(2603), 32
(2007); Z. Merali, Nature (London) {\bf 454}, 8 (2008); C. Bruder
and D. Loss, Physics {\bf 1}, 34 (2008).

\bibitem{scully}
M. O. Scully and K. Dr\"uhl, Phys. Rev. A {\bf 25}, 2208 (1982).

\bibitem{peres}
A. Peres, {\it Quantum Theory: Concepts and Methods}, (Kluwer Academic Publishers, Dordrecht, 1995), p.~373.

\bibitem{Katz2}
N. Katz, M. Neeley, M. Ansmann, R. C. Bialczak, M. Hofheinz, E.
Lucero,  A. O'Connell, H. Wang, A. N. Cleland, J. M. Martinis and A.
N. Korotkov, Phys. Rev. Lett. {\bf 101}, 200401 (2008).

\bibitem{restore-known} One of the possible procedures is the
    following. If the post-measured state is not pure, we can apply
    more measurements to make it pure, and then probabilistically apply
    a unitary operation from an easily calculable
    set, which creates a mixture of pure states identical to the
    initial state.

\bibitem{exception} An exception is when the initial state belongs
    to a very limited set, in which the measurement result
    corresponds to only one state.

\bibitem{Nielsen} M. A. Nielsen and I. L. Chuang, {\it Quantum computation
        and quantum information} (Cambridge University Press, Cambridge,
        2000).

\bibitem{Gardiner} C. W. Gardiner and P. Zoller, {\it Quantum noise}
    (Springer, Berlin, 2004), Ch. 2.2.3.

\bibitem{Ruskov-07}  R. Ruskov, A. Mizel, and A. N. Korotkov, Phys.
    Rev. B {\bf 75}, 220501(R) (2007).

\bibitem{note}
In this example, Plato could change his undoing strategy to simply
apply a tailored unitary to shift the disturbed state
$|\psi_{1,m}\ra$ back to its original state $|\psi_{1}\ra$.  Under
this modified strategy, the success probability will be ${\cal P}$,
which may or may not exceed $P_S$, depending on the strength of the
measurement.   However, in this case, Plato himself will not know
whether the strategy succeeded or not, and therefore cannot claim to
have an uncollapse procedure.

\bibitem{Jaynes}
E. T. Jaynes, {\it Probability Theory: The Logic of Science},
(Cambridge University Press, 2003).

\bibitem{Feller} W. Feller, {\it An Introduction to Probability Theory and
    Its Applications}, v. 1 (Wiley, NY, 1968).

\bibitem{Royer}
A. Royer, Phys. Rev. Lett. {\bf 73}, 913 (1994); {\bf 74},
    1040(E) (1995).

\bibitem{DDth}
S. A. Gurvitz, Phys. Rev. B {\bf 56}, 15215 (1997); A. N. Korotkov
and D. V. Averin, Phys. Rev. B {\bf 64}, 165310 (2001); H. S. Goan
and G. J. Milburn, Phys. Rev. B {\bf 64}, 235307 (2001); S. Pilgram
and M. B\"uttiker, Phys. Rev. Lett. {\bf 89}, 200401 (2002); A. A.
Clerk, S. M. Girvin, and A.~D. Stone, Phys. Rev. B {\bf 67}, 165324
(2003); D. V. Averin, in ``Exploring the Quantum-Classical Frontier:
Recent Advances in Macroscopic and Mesoscopic Quantum Phenomena'',
Eds. J. R. Friedman and S. Han (Nova Science, Huntington, NY, 2003);
cond-mat/0004364; A. Shnirman, D. Mozyrsky, and I. Martin, Europhys.
Lett. {\bf 67}, 840 (2004); A. A. Clerk and A. D. Stone, Phys. Rev.
B {\bf 69}, 245303 (2004); A. N. Jordan and M. B\"uttiker, Phys.
Rev. Lett. {\bf 95}, 220401 (2005); N. P. Oxtoby, P. Warszawski, H.
M. Wiseman, H. B. Sun, R.E.S. Polkinghorne, Phys Rev B {\bf 71},
165317 (2005); A. A. Clerk, M. H. Devoret, S. M. Girvin, F.
Marquardt,  and R. J. Schoelkopf, arXiv:0810.4729v1.

\bibitem{DDexp}
T. Hayashi, T. Fujisawa, H.-Du Cheong, Y.-Ha Jeong, and Y. Hirayama,
Phys. Rev. Lett. {\bf 91}, 226804 (2003); J.R. Petta, A.C.
Johnson, C.M. Marcus, M.P. Hanson, and A.C. Gossard, Phys. Rev.
Lett. {\bf 93}, 186802 (2004);  J.R. Petta,
A.C. Johnson, J.M. Taylor, E.A. Laird, A. Yacoby, M.D. Lukin, C.M.
Marcus, M. P. Hanson, and A.C. Gossard, Science {\bf 309}, 2180
(2005); J. M. Elzerman, R. Hanson, J. S. Greidanus, L. H. Willems
van Beveren, S. De Franceschi, L. M. K. Vandersypen, S. Tarucha,
and L. P. Kouwenhoven Phys. Rev. B {\bf 67}, 161308(R) (2003); A.
K. H\"uttel, S. Ludwig, K. Eberl, J. P. Kotthaus, Phys. Rev. B
{\bf 72}, R081310 (2005); J. Gorman, E. G. Emiroglu, D.G. Hasko, and D.A.
Williams, Phys. Rev. Lett. {\bf 95}, 090502 (2005).

\bibitem{S-normalization} We use normalization of the shot noise, in which
    $S_{I}=2eI(1-{\cal T})$, where ${\cal T}$ is the QPC transparency.


\bibitem{Kor-99} A. N. Korotkov, Phys. Rev. B {\bf 60}, 5737 (1999);
{\bf 63}, 115403 (2001); {\bf 67}, 235408 (2003).

\bibitem{Oksendal} B. {\O}ksendal, {\it Stochastic differential
        equations\/} (Springer, Berlin, 1998).

\bibitem{Jordan} A. N. Jordan and M. B\"uttiker, Phys. Rev. B {\bf 71}, 125333 (2005);
A. N. Jordan, A. N. Korotkov, and M. B\"uttiker, Phys. Rev. Lett.
{\bf 97},  026805 (2006); N. S. Williams and A. N. Jordan, Phys.
Rev. Lett. {\bf 100}, 026804 (2008).

\bibitem{ourprb}
A. N. Jordan and A. N. Korotkov, Phys. Rev. B {\bf 74}, 085307 (2006).

\bibitem{redner}  S. Redner, {\it A Guide to First-Passage Processes}
(Cambridge University Press, New York, 2001).

\bibitem{williams}
 N. S. Williams and A. N. Jordan, Phys. Rev. A {\bf 78}, 062322 (2008).

\bibitem{Nazarov-discuss}  It is easy to generalize Eqs.\
(\ref{B-l-rho})--(\ref{B-l-3}) to include the detector non-ideality
and cross-correlation between the output and back-action noises (as
in Ref.~[\onlinecite{Kor-99}]). Non-ideality leads to the extra
dephasing term $-\gamma_d\sigma_{12}$ in Eq.\    (\ref{B-l-3}) for
$\dot\sigma_{12}$, while the cross-correlation brings the term $i
K[I(t)-I_0]\sigma_{12}$ into the same equation. The formulation of
Eqs.\ (\ref{B-l-rho})--(\ref{B-l-3}) corresponds to the approach
developed in Ref.\ [\onlinecite{Wei-Nazarov}].

\bibitem{Wei-Nazarov} H. Wei and Yu. V. Nazarov, Phys. Rev. B {\bf 78}, 045308 (2008).

\bibitem{Martinis} J. M. Martinis, S. Nam, J. Aumentado, and C. Urbina,
Phys. Rev. Lett. {\bf 89}, 117901 (2002); K. B. Cooper, M. Steffen,
R. McDermott, R. W. Simmonds, S. Oh, D. A. Hite, D. P. Pappas, and
J.M. Martinis, Phys. Rev. Lett. {\bf 93}, 180401 (2004).

\bibitem{Katz} N. Katz, M. Ansmann, Radoslaw C. Bialczak, E. Lucero,
R. McDermott, M. Neeley, M. Steffen, E. M. Weig, A. N. Cleland, J.
M. Martinis, and A. N. Korotkov, Science {\bf 312}, 1498 (2006)

\bibitem{Dalibard} J. Dalibard, Y. Castin, and K. M{\o}lmer, Phys. Rev. Lett. {\bf
    68}, 580 (1992).

\bibitem{Pryadko} L. P. Pryadko and A. N. Korotkov, Phys. Rev. B {\bf
76}, 100503(R) (2007).

\bibitem{QPT-note} Strictly speaking, the QPT language may not be
applicable to the uncollapsing experiment, because it requires
linearity of the quantum operation, while after selection of
particular realizations and state normalization the quantum
operation is not necessarily linear. However, the linearity is
preserved when the selection probability does not depend on the
initial state, which is exactly the case for a perfect uncollapsing
-- see Eq.\ (\ref{tildeP}). It is also possible to show that even in
the imperfect non-linear case the uncollapsing fidelity defined via
the ``naive'' QPT language practically coincides with the rigorous
definition via the average state fidelity.\protect\cite{Keane}

\bibitem{Keane} K. Keane and A. N. Korotkov, in preparation.

\bibitem{Kor-ent} A. N. Korotkov, Phys. Rev A {\bf 65}, 052304 (2002);

\bibitem{spinwm}
A. N. Jordan, B. Trauzettel, and G. Burkard,
Phys. Rev. B {\bf 76}, 155324 (2007).

\bibitem{leo}
F. H. L. Koppens, C. Buizert, K. J. Tielrooij, I. T. Vink, K. C. Nowack,
T. Meunier, L. P. Kouwenhoven and L. M. K. Vandersypen, 
Nature {\bf 442}, 766 (2006).


\bibitem{elzerman}
J. Elzerman, R. Hanson, L. H. Willems van Beveren, B. Witkamp,  L.
M. K. Vandersypen, L. P. Kouwenhoven, Nature {\bf 430}, 431 (2004).

\bibitem{hanson}
R. Hanson, L. H. Willems van Beveren, I. T. Vink, J. M. Elzerman,
W. J. M. Naber, F. H. L. Koppens, L. P. Kouwenhoven, and L. M. K.
Vandersypen, Phys. Rev. Lett. {\bf 94}, 196802 (2005).



\bibitem{nuclear_exp}
A. C. Johnson, J. R. Petta, J. M. Taylor, A. Yacoby, M. D. Lukin,
C. M. Marcus, M. P. Hanson, A. C. Gossard, Nature {\bf 435}, 925
(2005); F. H. L. Koppens, J. A. Folk, J. M. Elzerman, R. Hanson, L.
H. Willems van Beveren, I. T. Vink, H. P. Tranitz, W. Wegscheider,
L. P. Kouwenhoven, and L. M. K. Vandersypen, Science {\bf 309}, 1346
(2005).

\bibitem{Ono}
K. Ono, D. G. Austing, Y. Tokura, and S. Tarucha, Science {\bf 297},
1313 (2002);   A. C. Johnson, J. R. Petta, C. M. Marcus, M. P.
Hanson, and A. C. Gossard, Phys. Rev. B {\bf 72}, 165308 (2005).

\bibitem{kim2009}
Y.-S. Kim, Y.-W. Cho, Y.-S. Ra, and Y.-H. Kim, arXiv:0903.3077.

\end{thebibliography}
\end{document}